\documentclass[a4paper,fleqn,usenatbib]{mnras}

\usepackage{natbib}
\usepackage{soul}

\usepackage{newtxtext,newtxmath}

\usepackage[T1]{fontenc}
\usepackage{ae,aecompl}

\usepackage{todonotes}
\usepackage{outlines}
\usepackage{color}

\usepackage{graphicx}	
\usepackage{amsmath}	
\usepackage{amssymb}	
\usepackage{bm}	        

\usepackage{natbib}

\newcommand{\dd}{\mathrm{d}}

\newcommand{\ms}{M_*}

\newcommand{\mpc}{\mathrm{Mpc}}
\newcommand{\hmpc}{h^{-1}\mathrm{Mpc}}

\newcommand{\hhmsol}{h^{-2}M_{\odot}}

\newcommand{\fgas}{f_{\mathrm{HI}}}
\newcommand{\dred}{\delta_{\mathrm{red}}}
\newcommand{\ssfr}{s{\mathrm{SFR}}}
\newcommand{\zgas}{Z_{\mathrm{gas}}}
\newcommand{\fdet}{f_{\mathrm{det}}}

\newcommand{\mgas}{M_{\mathrm{HI}}}

\newcommand{\sig}{\sigma_{\lg\,f_{\mathrm{HI}}}}

\newcommand{\kms}{\mathrm{km}\,s^{-1}}

\newcommand{\rom}[1]{\uppercase\expandafter{\romannumeral #1\relax}}

\title[Atomic gas in star-forming galaxies]{HI gas content of SDSS galaxies revealed by ALFALFA: implications
    for the mass-metallicity relation and the environmental dependence of HI in the local Universe}
\author[Zu 2018]{
Ying  Zu$^{1}$\thanks{E-mail: yingzu@sjtu.edu.cn},
\\
$^{1}$Department of Astronomy, Shanghai Jiao Tong University, Shanghai 200240, China
}

\date{Accepted XXX. Received YYY; in original form ZZZ}

\pubyear{2018}

\begin{document}

\label{firstpage}
\pagerange{\pageref{firstpage}--\pageref{lastpage}}
\maketitle

\begin{abstract}
    The neutral hydrogen~(HI) gas is an important barometer of recent star
    formation and metal enrichment activities in galaxies.  I develop a novel
    statistical method for predicting the HI-to-stellar mass ratio $\fgas$ of
    galaxies from their stellar mass and optical colour, and apply it to a
    volume-limited galaxy sample jointly observed by the Sloan Digital Sky
    Survey and the Arecibo Legacy Fast ALFA survey. I eliminate the impact of
    the Malmquist bias against HI-deficient systems on the $\fgas$ predictor by
    properly accounting for the HI detection probability of each galaxy in the
    analysis. The best-fitting $\fgas$ predictor, with an estimated scatter of
    $0.272$ dex, provides excellent description to the observed HI mass
    function. After defining an HI excess parameter as the deviation of the
    observed $\fgas$ from the expected value, I confirm that there exists a
    strong secondary dependence of the mass-metallicity relation on HI excess.
    By further examining the 2D metallicity distribution on the specific star
    formation rate vs. HI excess plane, I show that the metallicity dependence
    on HI is likely more fundamental than that on specific star formation rate.
    In addition, I find that the environmental dependence of HI in the local
    Universe can be effectively described by the cross-correlation coefficient
    between HI excess and the red galaxy overdensity $\rho_{cc}{=}-0.18$. This
    weak anti-correlation also successfully explains the observed dependence of
    HI clustering on $\fgas$. My method provides a useful framework for
    learning HI gas evolution from the synergy between future HI and optical
    galaxy surveys.
\end{abstract}




\vspace{1in}
\section{Introduction}
\label{sec:intro}

The neutral hydrogen~(HI) gas represents a key intermediate stage in baryon
cycling, between the initial accretion from the diffuse circumgalactic or
intergalactic medium~\citep{sancisi2008, tumlinson2017} and the formation of
dense molecular clouds that directly fuel star formation~\citep{kennicutt2012,
lada2012, leroy2013}. The variation of the HI gas reservoir usually precedes the
colour transformation of galaxies induced by star formation and
quenching~\citep{baldry2004, faber2007}, while regulating the metallicity of the
interstellar medium~(ISM) together with galactic outflows~\citep{dalcanton2007,
matteucci2012}. In this paper, I develop a statistical framework for connecting
the HI gas mass detected by ALFALFA~\citep{haynes2011} to the stellar mass and
optical colours of galaxies observed in SDSS~\citep{york2000}, and explore the
physical drivers of gas-phase metallicity and the environmental dependence of HI
within this framework.

As the most important measure of the HI content of a galaxy, the HI-to-stellar
mass ratio $\fgas$~(hereafter referred to as HI fraction) has been found to
correlate with the optical colour with a scatter of ${\sim}0.4$
dex~\citep{kannappan2004}. Subsequently, \citet{zhang2009} built a photometric
estimator of $\fgas$ by introducing an additional scaling of $\fgas$ with the
$i$-band surface brightness, reducing the scatter to $0.31$ dex. \citet{li2012}
later extended the $\fgas$ estimator by using a linear combination of four
parameters~(including stellar mass, stellar surface mass density $\mu_*$,
NUV-$r$ colour, and the $g{-}i$ colour gradient), resulting a slightly improved
scatter of $0.3$ dex and a more accurate match to the high-$\fgas$ systems
observed by ALFALFA. Alternatively, non-linear predictors have been recently
developed using machine learning algorithms, which usually require training over
a large number of HI-detected systems~\citep{teimoorinia2017, rafieferantsoa2018}, or
simple functional fits to the median $\fgas$ trend with stellar
mass~\citep{maddox2015, calette2019}.

However, current HI surveys like ALFALFA are relatively shallow in depth, and
are thus systematically biased against low-$\fgas$ systems at any given
redshift. Consequently, any $\fgas$ predictor inferred or trained exclusively
from systems above the HI detection threshold would be plagued by the Malmquist
bias, overestimating the $\fgas$ for systems that are missed by the HI survey.
Such Malmquist bias can be partially alleviated by observing a smaller volume to
a higher depth in HI. For example, using a roughly $\fgas$-limited but
significantly smaller sample~(GASS; GALEX Arecibo SDSS Survey),
\citet{catinella2010} constructed a $\fgas$ predictor using the linear
combination of NUV-$r$ colour and $\mu$, resulting in a scatter of ${\sim}0.3$
dex~\citep[see also][]{catinella2013}. Without having to trade volume for depth,
we develop a new method to eliminate the Malmquist bias when predicting $\fgas$
from the stellar mass and colour of SDSS galaxies, by properly accounting for
the ALFALFA detection probability of each SDSS galaxy in the analysis.

Beyond $\fgas$, the metal abundance within the gas serves as the fossil record
of the chemical enrichment history, reflecting the complex interplay between
star formation and gas accretion during the baryon cycling~\citep{peeples2014}.
For star-forming galaxies, gas-phase metallicity is tightly corrected with
stellar mass in the oxygen-to-hydrogen abundance ratio, forming the well-known
mass-metallicity relation~\citep[MZR;][]{tremonti2004}. It has been suggested
that the star formation rate~(SFR) could drive the scatter in MZR.
\citet{ellison2008} first reported the existence of a secondary dependence of
the MZR on SFR, and \citet{mannucci2010} later proposed that galaxies observed
up to z$\sim$2.5 define a tight surface in the 3D space of stellar mass, SFR,
and gas-phase metallicity~(a.k.a., the fundamental metallicity relation; FMR),
with a residual dispersion of $0.05$ dex in metallicity~\citep[see
also][]{laralopez2010, andrews2013}.

The existence of this secondary dependence, however, has been questioned by
studies based on the integral field spectroscopy~\citep[IFS;][]{sanchez2013,
barreraballesteros2017, sanchez2017, sanchez2019}, contrary to many results
based on single-aperture spectroscopic surveys\citep[e.g., SDSS]{yates2012,
salim2014, cresci2019}. In particular, \citet{sanchez2013} claimed that the
secondary relation could be explained by a pure aperture effect in the SDSS
survey. Large discrepancies also persist among various studies based on the
single-aperture data, which may arise from the use of different metallicity
indicators~\citep{kashino2016} and different galaxy selection criteria adopted
in different studies~\citep{salim2014, telford2016}.  Nevertheless, various
theoretical models have subsequently been proposed to explain the MZR, assuming
SFR is the main process that shaped the MZR~\citep{peeples2011, dave2012,
dayal2013, lilly2013, zahid2014}.

Besides star formation, it is reasonable to expect that gas accretion plays a
role in regulating the metallicity of the ISM. Indeed, \citet{bothwell2013}
showed that the MZR of ${\sim}4000$ ALFALFA galaxies exhibits a strong secondary
dependence on HI mass, with HI-rich galaxies being more metal poor at fixed
stellar mass. Using 260 nearby galaxies from the {\it Herschel} Reference
Survey, \cite{hughes2013} detected a similar anti-correlation between gas
fraction and oxygen abundance at fixed stellar mass, but almost no environmental
dependence of the MZR.  Applying a principal component analysis over ${\sim}200$
galaxies compiled from several molecular gas surveys, \citet{bothwell2016a}
further argued that the underlying driver of MZR is the molecular gas mass, and
the FMR is merely a by-product of molecular FMR via the Kennicutt-Schmidt
law~\citep{bothwell2016b}. More recently, by stacking the HI spectra of
star-forming galaxies along the MZR, \citet{brown2018} confirmed the strong
anti-correlation between HI mass and gas-phase metallicity at fixed stellar
mass, providing further evidence that the scatter in the MZR is primarily driven
by fluctuations in gas accretion. To ascertain whether SFR or HI mass is the
more fundamental driver, I will present a comprehensive analysis of
metallicity, SFR, and HI mass for a large sample of galaxies jointly observed by
SDSS and ALFALFA.

In addition to the optical properties of each galaxy, the HI gas reservoir also
depends on the large-scale density environment. For example, it is long known
that satellite galaxies in massive halos are deficient in HI~\citep{haynes1984,
boselli2006, yoon2015, jaffe2015}, due to processes like the ram-pressure and
tidal stripping~\citep{gunn1972, merritt1983, moore1996, abadi1999,
mccarthy2008, kronberger2008, bekki2009}. Gas accretion history may be tied to
the halo growth history, which is known to be correlated with the large-scale
environment~\citep{fakhouri2010}. The environmental dependence of cosmic HI
distribution can be predicted using semi-analytic models~\citep{fu2010, xie2018}
and hydro-dynamic simulations~\citep{dave2017}, or statistically accounted for
within the halo model~\citep{guo2017, obuljen2018}. However, a quantitative
description of the environmental dependence of HI is still lacking. In my
analysis, I quantify this dependence using the cross-correlation coefficient
$\rho_{cc}$ between HI excess and galaxy overdensity, and develop three
independent approaches to measuring $\rho_{cc}$ directly from data.

This paper is organized as follows. I briefly describe the data and the joint
SDSS-ALFALFA sample in \ S~\ref{sec:data}, and introduce my likelihood model in
\S~\ref{sec:method}.  I present my main findings on the mass-metallicity
relation in \S~\ref{sec:mzr} and the environmental dependence of HI in
\S~\ref{sec:env}. I conclude by summarizing my results and looking to the
future in \S~\ref{sec:conc}.

Throughout this paper, I assume the {\it WMAP9} cosmology~\citep{wmap2013} for
distance calculations. All the length and mass units in this paper are scaled as
if the Hubble constant were $100\,\kms\mpc^{-1}$. In particular, all the
separations are co-moving distances in units of $\hmpc$, and the stellar and HI
mass are both in units of $\hhmsol$. I use $\lg x{=}\log_{10} x$ for the
base-$10$ logarithm.

\section{Data}
\label{sec:data}
\subsection{SDSS Volume-Limited Stellar Mass Sample}
\label{subsec:sdss}

I make use of the final data release of the Sloan Digital Sky
Survey~\citep[SDSS DR7;][]{york2000, abazajian2009}, which contains the
completed data set of the SDSS-I and the SDSS-II. In particular, I obtain the
Main Galaxy Sample~(MGS) data from the \texttt{dr72} large--scale structure
sample \texttt{bright0} of the ``New York University Value Added
Catalogue''~(NYU--VAGC), constructed as described in~\citet{blanton2005}.  The
\texttt{bright0} sample includes galaxies with $10{<}m_r{<}17.6$, where $m_r$ is
the $r$-band Petrosian apparent magnitude, corrected for
Galactic extinction. I apply the ``nearest-neighbour'' scheme to correct for the $7\%$ galaxies that are
without redshift due to fibre collision, and use data exclusively within the contiguous area in the North
Galactic Cap and regions with angular completeness greater than $0.8$.

I employ the stellar mass and gas-phase metallicity estimates from the latest
MPA/JHU value-added galaxy
catalogue\footnote{\url{http://home.strw.leidenuniv.nl/~jarle/SDSS/}}. The
stellar masses were estimated based on fits to the SDSS photometry following the
philosophy of~\citet{kauffmann2003} and~\citet{salim2007}, and assuming the
Chabrier~\citep{chabrier2003} initial mass function~(IMF) and
the~\citet{bruzual2003} SPS model.  The MPA/JHU stellar mass catalogue is then
matched to the NYU-VAGC \texttt{bright0} sample.  For the gas-phase
metallicities of galaxies, I adopt the Bayesian metallicity estimates from
fitting to multiple strong nebula emission lines~\citep{tremonti2004}.

From the \texttt{bright0} catalogue, I select a volume-limited sample of
$14{,}140$ galaxies with log-stellar mass $\lg\,(\ms/\hhmsol)\geq 9.4$ and
redshift range $z\in [0.016, 0.04]$, which forms the basis sample for my joint
analysis with ALFALFA~(as will be described below). Although the ALFALFA survey
robustly detected HI sources up to $z{\sim}0.05$ before encountering the radio
frequency interference~(RFI), I choose a slightly lower redshift limit of
$z_{\mathrm{max}}{=}0.04$ so that the volume-limited sample can reach a lower
threshold in stellar mass, hence a higher fraction of HI detections within the
SDSS sample.

\begin{figure}
\begin{center}
    \includegraphics[width=0.48\textwidth]{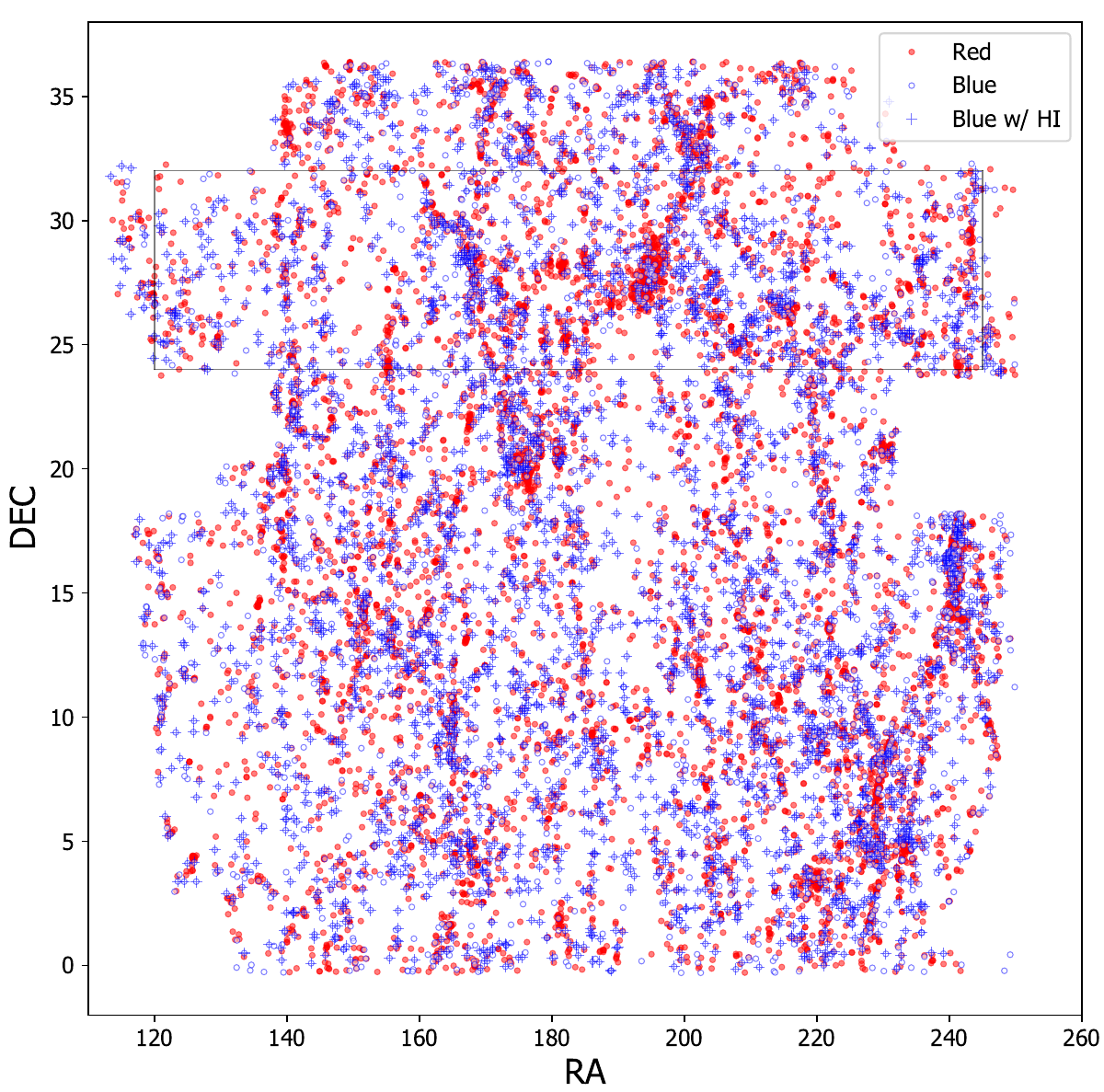} \caption{The
    overlapping footprint between SDSS and ALFALFA in the northern galactic cap.
    Red filled and blue open circles indicate the red and blue galaxies observed
    by SDSS, respectively. Blue galaxies that are detected in ALFALFA are
    additionally marked by blue crosses. The SDSS sample is volume-limited with
    $\lg\,(M_*/\hhmsol)\geq 9.4$
	and $z\in [0.016,\,0.04]$. The region enclosed by the rectangular box is
	further highlighted in Figure~\ref{fig:wedge}.}
\label{fig:footprint}
\end{center}
\end{figure}

\subsection{The ALFALFA $\mathbf{\alpha.100}$ HI Sample}
\label{subsec:alpha}

The Arecibo Fast Legacy ALFA~\citep[ALFALFA;][]{haynes2011} survey is a blind
extragalactic HI survey conducted using the seven-horn Arecibo L-band Feed
Array~(ALFA) onboard the $305$-m Arecibo telescope.  In order to reveal the
faint-end population of the HI mass function in the local Universe~($z{<}0.05$),
ALFALFA mapped ${\sim}7000$ deg$^2$ of two contiguous high Galactic latitude
regions between $2005$ and $2011$, searching for HI line emission across the
entire frequency range between $1335$ and $1435$ $MHz$. I make use of the final
data
release~\citep[$\alpha.100$\footnote{\url{http://egg.astro.cornell.edu/alfalfa/}};][]{Haynes2018},
which contains ${\sim}31500$ sources up to $z=0.06$.  Due to the minimal overlap
between the ALFALFA and SDSS footprints in the southern Galactic cap, I will
focus exclusively on the northern Galactic region in my joint analysis of the
two surveys.

Each HI detection is characterised by its angular position on the sky, radial velocity, velocity width
$W_{50}$, and integrated HI line flux density $S_{21}$. The HI mass of each system can be estimated as
\begin{equation}
    \mgas = 2.356\times 10^5\,D^2\,S_{21}\; [\hhmsol],
\end{equation}
where $D$ is the distance to the source measured in units of $\hmpc$, and
$S_{21}$ in units of $\mathrm{Jy}\,\kms$. Each detection is then assigned a
detection category code depending on several reliability indicators, including
the signal-to-noise ratio~(S/N) of the detection and whether there exists an
optical counterpart identified in other surveys~(mainly SDSS). In particular,
the ``Code 1'' sources~($25434$) are reliable detections with S/N above $6.5$,
while a subset of those below $6.5$ are assigned ``Code 2''~($6068$) due to
having identified optical counterparts. In general, the Code 2 detections are
also highly reliable despite having a relatively low S/N. Therefore, I will
include both categories of detections in my analysis, and develop a likelihood
model to self-consistently account for the non-detections.

\subsection{Cross-matched SDSS and ALFALFA Sample}
\label{subsec:matched}

To study the HI content of optically selected galaxies, I cross-match the SDSS
\texttt{bright0} and the ALFALFA $\alpha.100$ catalogues across their shared
footprint in the northern Galactic cap. For each SDSS
galaxy I first find all its potential HI counterparts by adopting a search radius of $36$ arcsec, $80\%$ larger
than the typical centroiding uncertainty of ALFALFA~\citep[${\sim}20$
arcsec;][]{haynes2011}. I flag it as an HI non-detection if no ALFALFA source
is found within $36$ arcsec of that galaxy; If the search returns one or
multiple HI candidates, I then examine if the radial velocity of each candidate
falls within $\pm 600 \,\kms$ of the SDSS redshift. To ensure that the match is
unique, I always choose the closest HI source (in 2D) when there are multiple
candidates left after the two passes.

After the cross-match, each SDSS galaxy in the volume-limited stellar mass
sample described in \S\ref{subsec:sdss} is either detected in ALFALFA with a
reliable HI mass, or an HI non-detection due to the lack of an HI emission with
S/N above $6.5$. For each SDSS galaxy detected in HI, I characterise the HI
richness of the system by defining an HI-to-stellar mass ratio
$\fgas{\equiv}\mgas/\ms$, which I refer to as {\it HI fraction} throughout the
rest of the paper. For the galaxies that are not detected in HI, I emphasize
that the resulting Malmquist bias not only has to be properly accounted for in
the analysis, but they should also provide important clues as to which kind of
SDSS galaxies~(in terms of $\ms$ and $g{-}r$) are intrinsically more likely to
be HI-deficient than those that are observed.

It is worth nothing that, the spectroscopic measurement of each SDSS galaxy is
confined within a $3$-arcsec aperture~(i.e., diameter of the fiber) centred on
that galaxy. To correct for this aperture effect on SFR, the {\it total} SFR of
galaxies are derived from empirically extrapolating the {\it fiber} SFR using
broad-band colours~\citep{brinchmann2004, salim2007}. However, there is no such
aperture correction available for metallicities and the ALFALFA observation has
a much larger effective aperture than SDSS. The mixing of different physical
scales due to the different observational apertures may introduce some spurious
correlations in my joint metallicity-SFR-HI analysis. This aperture bias can be
alleviated by using the same aperture for metallicity and SFR measurements with
resolved spectroscopy from IFS observations, as done
by~\citet{barreraballesteros2018}. However, due to the lack of real HI
observations with matching aperture, they had to derive gas fraction from dust
extinction inferred from the optical spectra, which could lead to spurious
correlations of gas fraction with both SFR and metallicity. Therefore, in order
to focus on the physical processes at the disk-halo interface, I employ the
aperture-corrected total SFR and the direct HI mass from ALFALFA for my
analysis, and use the fiber metallicity to study the implications of those
processes for the chemical enrichment in the central region of galaxies.  I
will return to the impact of aperture bias later in \S~\ref{sec:mzr}.

\begin{figure*}
\begin{center}
    \includegraphics[width=0.96\textwidth]{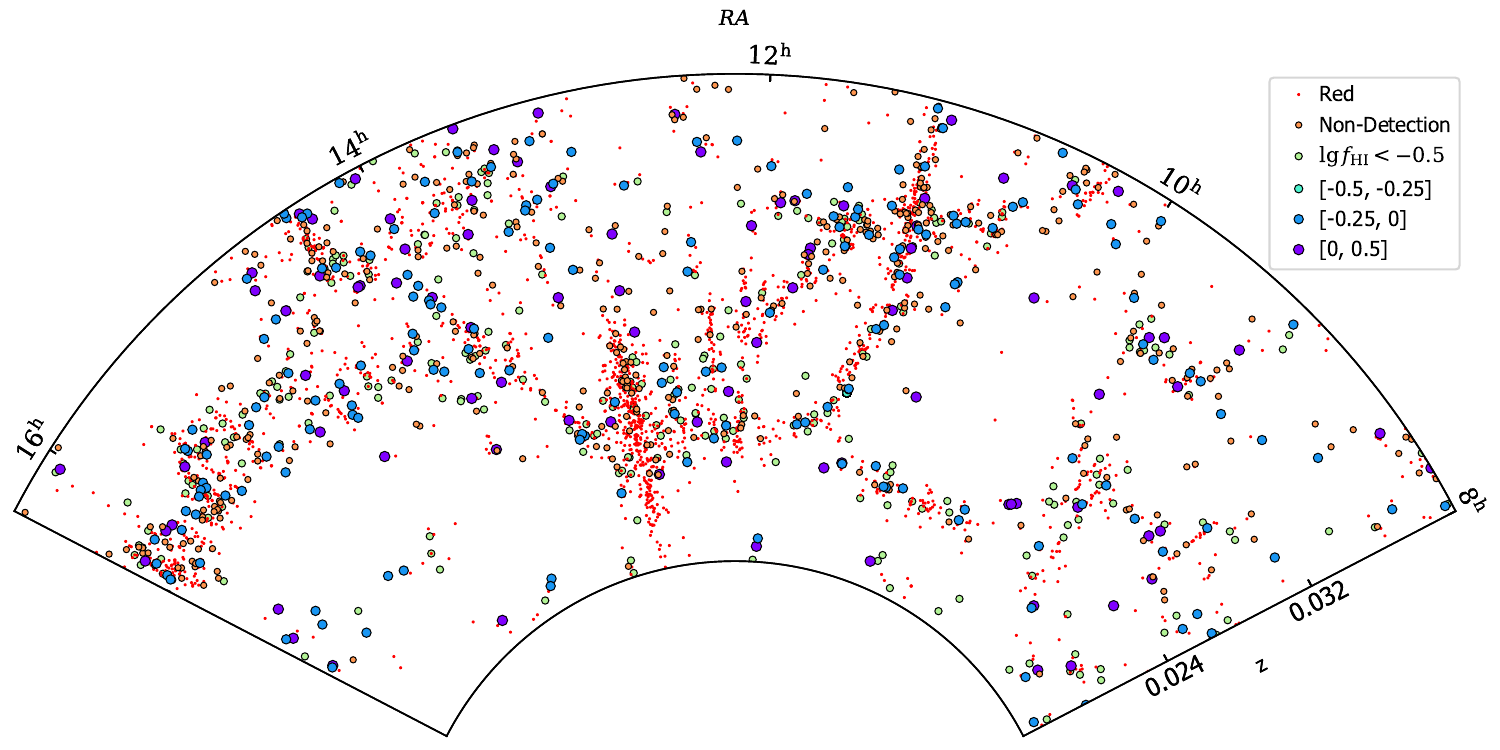} \caption{Redshift
    distribution of galaxies in the SDSS-ALFALFA joint
    sample~($120{<}\mathrm{RA}{<}245$ and $24{<}\mathrm{DEC}{<}32$; highlighted
    by the rectangular box in Figure~\ref{fig:footprint}). Red dots
	and coloured circles indicate the positions of the red and blue SDSS
	galaxies, respectively. Different sizes and colours of the circles
	correspond to the five levels of HI gas fraction $\fgas$ observed by
	ALFALFA~(from ``non-detection'' to $\fgas{>}1$), indicated by the legend
	on the top right. The dominant structure at $z{\sim}0.023$ and
	$\mathrm{RA}{\sim}13^h$ is the \texttt{Coma} cluster.}
\label{fig:wedge}
\end{center}
\end{figure*}

\begin{figure*}
\begin{center}
    \includegraphics[width=0.96\textwidth]{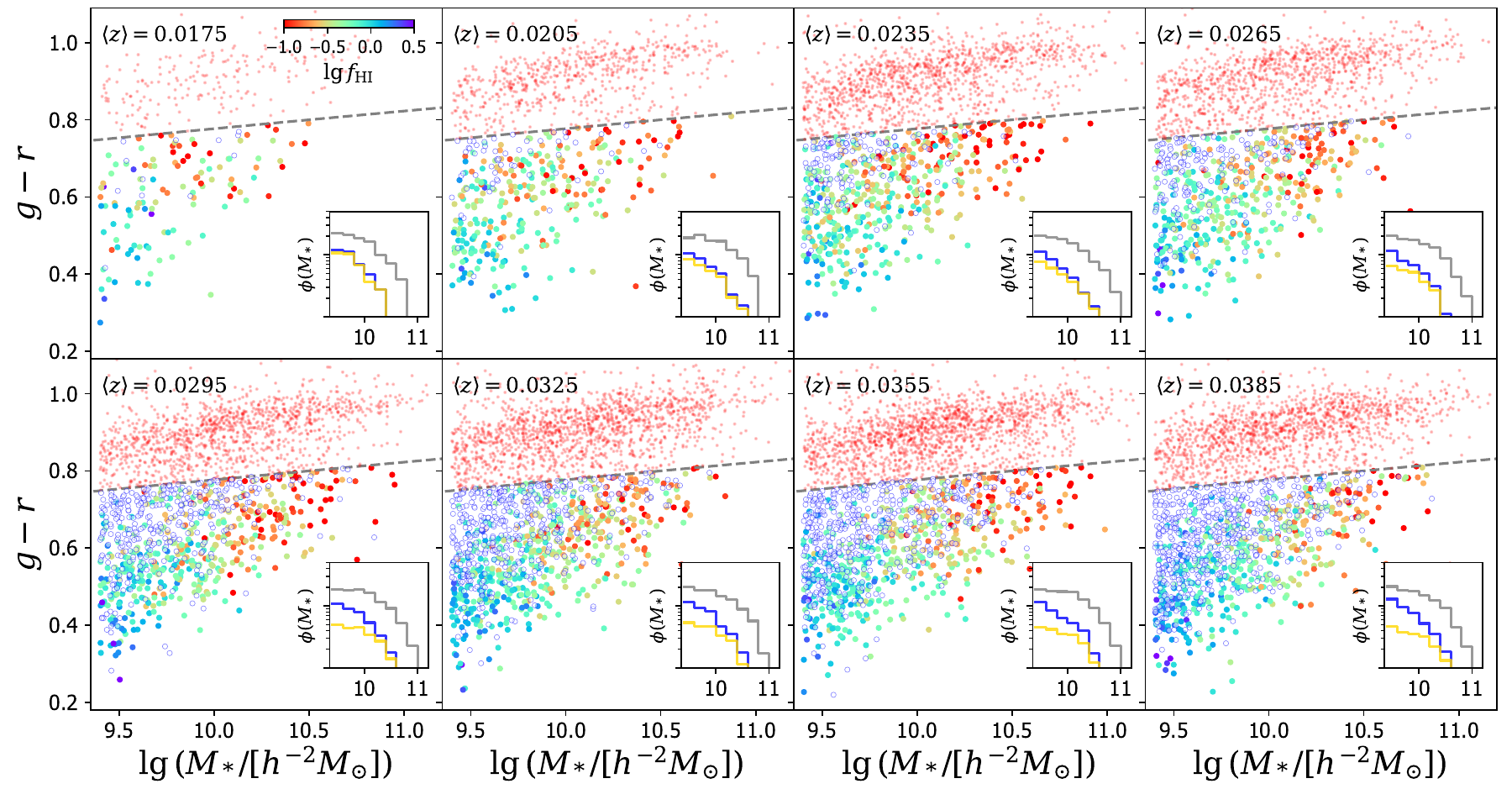} \caption{Colour-mass diagrams of the
	SDSS-ALFALFA joint sample at eight different redshift slices~(between $0.016$ and $0.04$ with $\Delta
	z {=}0.003$). In each panel, red dots indicate the red galaxies, defined as those above the colour
	cut~(gray dashed line) on the diagram. Below the colour cut, blue open circles indicate the blue
	galaxies that are not detected in ALFALFA, while the colour-filled circles are the blue galaxies
	detected with different HI gas fractions $\fgas$, colour-coded by the colour bar in the top left
	panel. The inset panel inside each panel shows the stellar mass functions of the total~(gray),
	blue~(blue), and ALFALFA-detected blue~(gold) galaxies.  While the stellar mass functions of the total
	and blue galaxies remain unchanged with redshift, the HI-detection completeness of ALFALFA decreases
	rapidly with increasing redshift.}
\label{fig:ms-color_matched}
\end{center}
\end{figure*}

I plan to focus on the star-forming population in my analysis, as the majority
of quenched galaxies are not detected in ALFALFA and those detected in HI do not
follow the same gas scaling relations as the star-forming
ones~\citep{boselli2014}. Therefore, I divide galaxies into quenched~(red) and
star-forming~(blue) based on their $g-r$ colours~({\it k}-corrected to
$z{=}0.1$). I use broad-band colours rather than the star formation rate~(SFR),
because I am interested in building a photometric estimator of HI fraction
that do not rely on high S/N spectroscopic observations.  For the same reason,
we did not remove the $1{,}650$ type 2 Active Galactic Nucleus~(AGN) candidates
from the blue sample using the BPT diagnostics, which rely on emission line
indices that usually require high S/N spectra. Following \citet{zu2016}, we
adopt a stellar mass-dependent colour cut to divide galaxies into red and blue,
\begin{equation}
\left(g-r\right)_{\mathrm{cut}}(\ms) = 0.8 \left(\frac{\lg\ms [\hhmsol]}{10.5}\right)^{0.6},
\label{eqn:cut}
\end{equation}
indicated by the gray dashed lines in Fig.~\ref{fig:ms-color_matched}~(described further below).

To summarize, my volume-limited SDSS-ALFALFA joint sample includes $8{,}721$
red and $5{,}419$ blue galaxies with stellar mass above $\lg\,(M_*/\hhmsol)=9.4$
and redshifts between $0.016$ and $0.04$.  I will focus exclusively on this
joint sample throughout the rest of the paper. Figure~\ref{fig:footprint} show
the
distribution of red~(red dots) and blue~(blue circles) galaxies of my joint sample across the shared footprint between SDSS and ALFALFA. Among the $5{,}419$ blue galaxies, $3{,}258$~($60\%$)
of them were detected in HI by ALFALFA~(blue crosses; including both Code 1 and
2 detections), and $2{,}161$ are non-detection in ALFALFA, respectively.  The
region enclosed by the gray rectangular box is further highlighted in
Figure~\ref{fig:wedge}, which indicates the redshift and RA distribution of red
galaxies~(red dots) and blue galaxies with five different levels of
$\fgas$~(colour-filled circles): HI non-detection~(orange),
$\lg\,\fgas{<}-0.5$~(green), $-0.5{\leq}\lg\,\fgas{<}-0.25$~(cyan),
$-0.25{\leq}\lg\,\fgas{<}0$~(blue), and $0{\leq}\lg\,\fgas{<}0.5$~(purple). The
\texttt{Coma} cluster can be clearly seen as the dominant structure at
$z{\sim}0.023$ and $\mathrm{RA}{\sim}13^h$.

Figure~\ref{fig:ms-color_matched} shows the colour-mass diagrams of eight
different redshift bins~($\Delta z{=}0.003$) between $z{=}0.016$ and $0.04$. In
each panel, the red dots above the gray dashed line~(Equation.~\ref{eqn:cut})
represent the quenched/red galaxies, while the colour-filled and blue open
circles below are blue galaxies with and without detection in HI, respectively.
The colour-coding of the filled circles indicate the value of $\lg\,\fgas$, as
described by the colourbar shown in the top left panel. The inset panels show
the observed stellar mass functions of total~(gray histograms), blue~(blue), and
HI-detected blue~(yellow) galaxies at respective redshifts. The total and blue
histograms stay roughly unchanged across the eight redshift bins, a manifest of
the high stellar mass-completeness of the SDSS volume-limited sample. However,
the yellow histograms decrease substantially at the low mass end towards higher
redshifts, signaling the strong Malmquist bias of the HI detection rate in
ALFALFA.

Within the blue population shown in Figure~\ref{fig:ms-color_matched}, the HI
non-detections mainly occupy the high-$g{-}r$ and low-$\ms$ corner of the
so-called ``blue cloud'' at low redshifts~(top panels), but spread out to the
entire cloud in the highest redshift bin~(bottom right panel). Among the
HI-detected galaxies, the HI fraction exhibits strong decreasing trends with
both $g{-}r$ and $\ms$ in all panels. The trend with $g{-}r$ is likely real,
because at fixed $\ms$ the non-detections are preferentially redder and have a
smaller average $\mgas$~(hence $\fgas$) than those detected in HI. However, it
is unclear whether the trend with $\ms$ seen in each panel is physical --- at
fixed colour the non-detections have on average lower $\ms$ and lower $\mgas$
than those detected in HI, but the two populations may have similar $\fgas$. In
the next Section, I will build a rigorous likelihood model to quantify the
level of intrinsic correlation between $\fgas$ and $\ms$ despite the obscuration
caused by the Malmquist bias.

\section{Methodology}
\label{sec:method}

\subsection{HI Fraction Predictor and Detection Probability Model}
\label{subsec:hifp}

Inspired by the two roughly independent trends of $\lg\fgas$ with $\lg\ms$ and $g{-}r$ seen in
Figure~\ref{fig:ms-color_matched}, I construct a linear mixture model for the HI fraction predictor~(HI-FP)
\begin{equation}
    \lg\fgas  = a \times \lg\ms + b \times (g{-}r) + c + \sig \times \epsilon,
    \label{eqn:fp}
\end{equation}
where $a$, $b$, and $c$ are the three parameters that determine $\langle\lg\fgas | \ms,g{-}r\rangle$~(i.e.,
the expected value of $\lg\fgas$ for any galaxy with given $\ms$ and $g{-}r$), while $\epsilon$ is a Gaussian
random variable with a zero mean and a unit variance.

\begin{figure}
\begin{center}
    \includegraphics[width=0.48\textwidth]{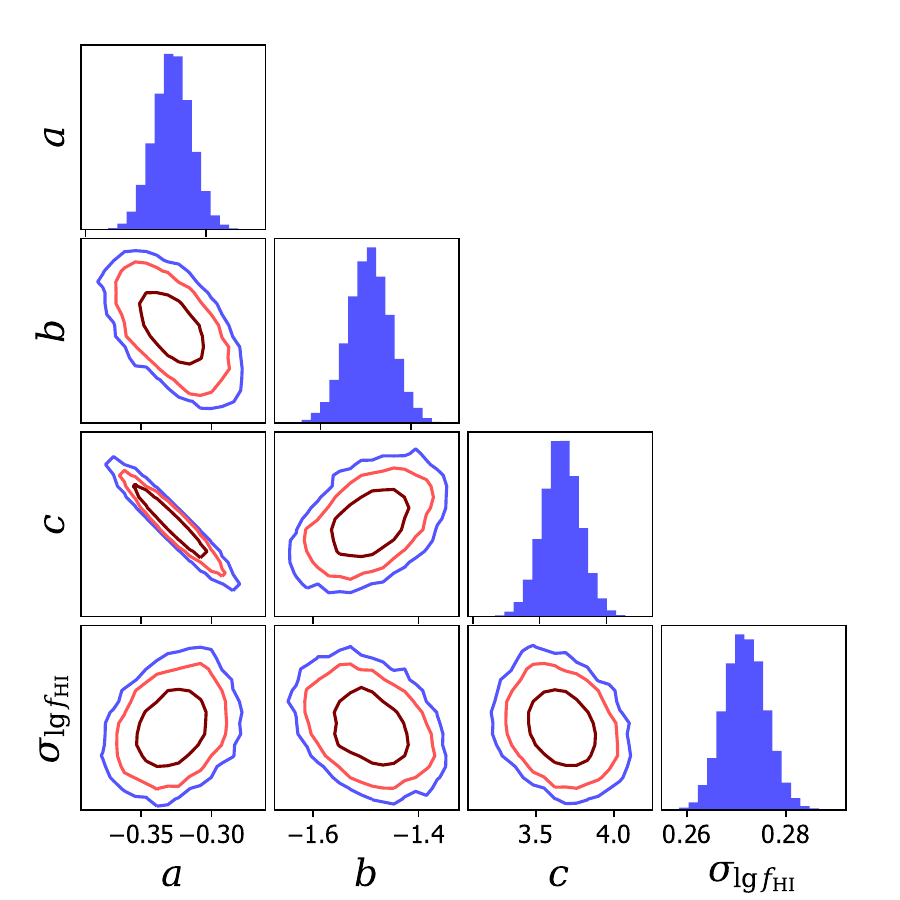} \caption{Parameter constraint for the HI gas
    fraction predictor: $\lg\,\fgas{=} a \lg\,\ms + b (g{-}r) + c + \epsilon \sigma_{\lg\fgas})$,
	marginalized over the six nuisance parameters that describe the detection probability as a function
        of HI mass and redshift. Red, magenta, and blue contours in the off-diagonal panels indicate the $68\%$,
	$95\%$, and $98\%$ confidence levels, respectively.}
    \label{fig:glory}
\end{center}
\end{figure}

Note that $\sig^2$ is the quadratic sum of the intrinsic scatter and the 1-$\sigma$ measurement uncertainty.
However, since the measurement error on $\fgas$ reported by ALFALFA is rather uniform across the
sample~(${\sim}0.05$ dex), I do not treat the two scatter components separately. In
addition, I assume a
constant log-normal scatter about the mean HI fraction at fixed $\ms$ and $g{-}r$. I have tried incorporating redshift and
stellar mass dependent scatters but failed to detect any such dependences within the data.

\begin{figure*}
\begin{center}
    \includegraphics[width=0.96\textwidth]{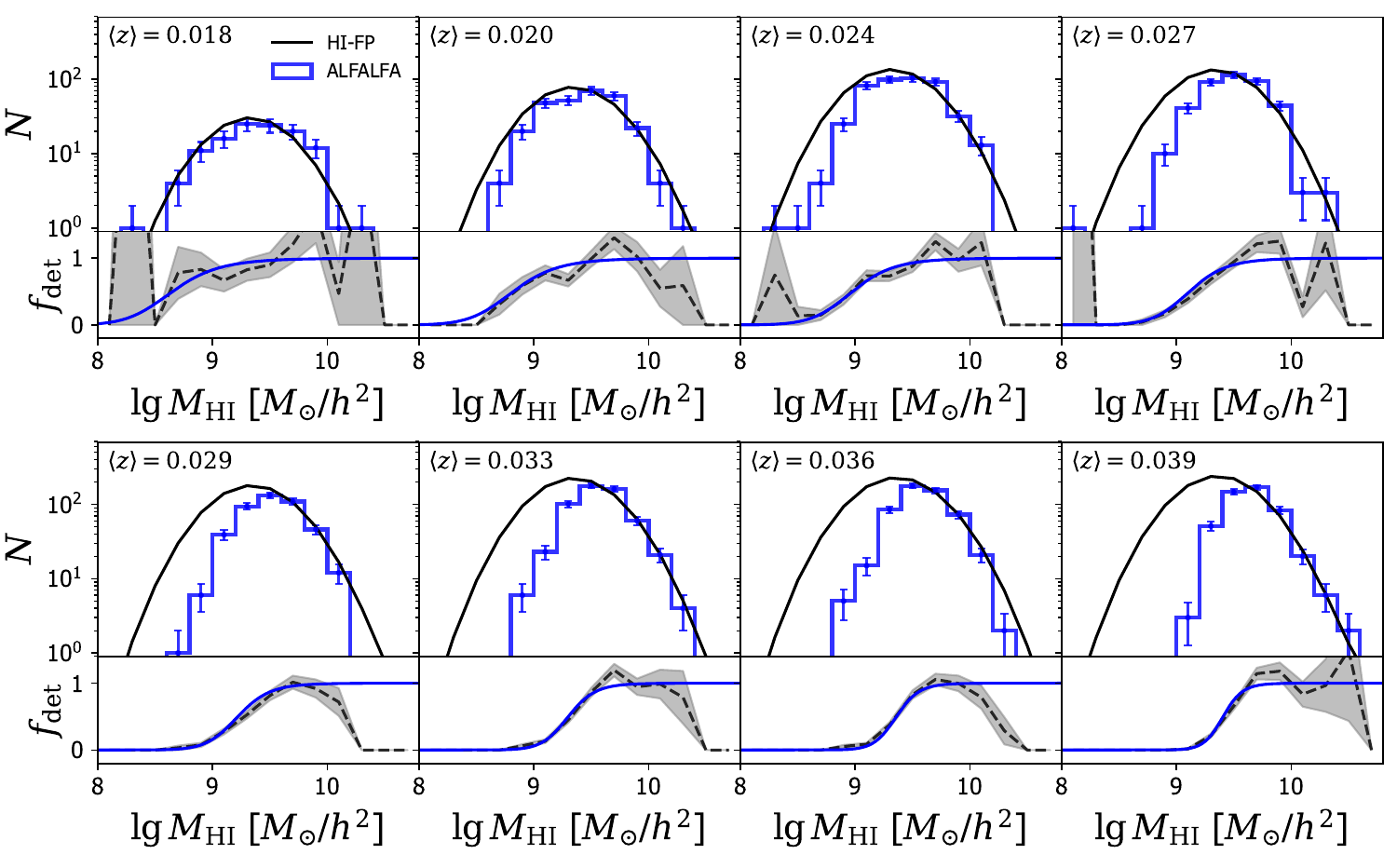} \caption{Comparison between the
	observed~(blue histograms) and predicted~(black curves) HI mass functions at eight different redshift
	slices~(top sub-panels).  In each bottom sub-panel, the black dashed curve shows the ratio between the
	observed and predicted HI mass functions, i.e., the HI-detection probability of ALFALFA in that
	redshift slice, with the gray shaded region indicating the Poisson uncertainties; The blue solid curve
	indicates the detection probability predicted from the best-fitting model.}
    \label{fig:detection_rate}
\end{center}
\end{figure*}

For the HI detection rate $\fdet$, \citet{haynes2011} demonstrated that it is mainly a function of $S_{21}$,
modulo some dependence on $W_{50}$~(see their Fig. 12) at fixed $S_{21}$, so that more extended HI emission
lines are less likely to be detected. However, at fixed redshift $z$, I find that the observed distribution
of $W_{50}$ at fixed $S_{21}$ only depends weakly on galaxy stellar mass and colour.
Therefore, I can simply
parameterise $\fdet$ as a function of HI mass $\mgas$,
\begin{equation}
  \fdet(\lg\mgas \mid z)=
  \begin{cases}
   {f(\lg\mgas \mid z)}\,/\,{f(12 \mid z)} & \text{if } \lg\mgas \leq 12 \\
   1       & \text{if } \lg\mgas > 12,
  \end{cases}
  \label{eqn:fdet}
\end{equation}
and
\begin{equation}
    f(\lg\mgas \mid z) = \frac{(\lg\mgas - 7)^{\mu_z}}{(\lg\mgas - 7)^{\mu_z} + (\lg M_{\mathrm{HI},z} -
    7)^{\mu_z}},
\end{equation}
where $\lg M_{\mathrm{HI},z}$ is the characteristic logarithmic HI mass at which the detection rate equals to
$50\%$ at redshift $z$, while $\mu_z$ controls the slope of the decline from $100\%$ at high $\mgas$ to $0\%$
at low $\mgas$. To compute $\lg M_{\mathrm{HI},z}$ and $\mu_z$ at arbitrary redshift $z$,
I choose three
pairs of ($\lg M_{\mathrm{HI},z}$, $\mu_z$) at $z{=}0.016,\,0.028,\,0.04$ as free parameters, and use the
\texttt{cubic} \texttt{spline}~\citep{nr} method to smoothly interpolate the two parameters of $\fdet(\lg\mgas
\mid z)$.

Combining the HI fraction predictor and the HI detection probability model, I now have ten parameters
$\bm{\theta}{\equiv}\{a,\,b, \,c, \,\sig, \,\lg M_{\mathrm{HI},0}, \,\mu_0, \,\lg M_{\mathrm{HI},1}, \,\mu_1,
\,\lg M_{\mathrm{HI},2}, \,\mu_2\}$, where the subscripts $0,\,1,\,2$ indicate the three pivot redshifts
$0.016,\,0.028,\mathrm{and}\,0.04$, respectively. Among the ten parameters, ${a,\,b, \,c,
\,\sig}$ are my key parameters
that describe the HI fraction predictor, while the rest are nuisance parameters for characterising the HI
incompleteness within the stellar mass-complete sample.

\subsection{Likelihood Model}
\label{subsec:likelihood}

My input data $\bm{\mathcal{D}}$ consist of $N{=}5{,}419$ blue galaxies in the SDSS-ALFALFA joint sample, each
observed with three features $\{\ms,\,g{-}r,\,z\}$. Among the $N$ galaxies, $n{=}3{,}258$ of them are observed
with reliable HI fraction $\fgas$, while the rest $N{-}n$ are non-detections. My goal is to derive the
posterior probability distribution function~(PDF) of $\bm{\theta}$ given $\bm{\mathcal{D}}$, $P(\bm{\theta}
\mid \bm{\mathcal{D}})$, i.e., the product of the likelihood function $P(\bm{\mathcal{D}}\mid \bm{\theta} )$
and the prior $P(\bm{\theta})$. I adopt flat priors on all the ten parameters in my analysis.

Armed with the models for the HI fraction predictor~(Equation.~\ref{eqn:fp}) and the HI
incompleteness~(Equation.~\ref{eqn:fdet}), I can derive the likelihood function $P(\bm{\mathcal{D}} \mid
\bm{\theta})$ analytically by decomposing it into two components,
\begin{equation}
    P(\bm{\mathcal{D}} \mid \bm{\theta}) = P(\bm{\mathcal{D}}_{\mathrm{det}} \mid \bm{\theta})
    \times P(\bm{\mathcal{D}}_{\mathrm{non-det}} \mid \bm{\theta}),
\end{equation}
where
\begin{equation}
    P(\bm{\mathcal{D}}_{\mathrm{det}} \mid \bm{\theta}) = \prod_{i=1}^n \; \fdet(\lg\mgas^i \mid
    z^i,\,\bm{\theta}) \; P(\lg\mgas^i\mid
    \ms^i, g^i{-}r^i, \bm{\theta})
\label{eqn:ldet}
\end{equation}
and
\begin{equation}
\begin{split}
    P(\bm{\mathcal{D}}_{\mathrm{non-det}} \mid \bm{\theta}) & = \prod_{j=1}^{N-n}
    \;\int_0^{\lg\mgas^{\mathrm{max}}} [(1 - \fdet(\lg\mgas^{\prime} \mid z^j,\,\bm{\theta})] \\
     &\;\;\;\;\times \; P(\lg\mgas^{\prime}\mid \ms^j, g^j{-}r^j, \bm{\theta})\;\dd\,\lg\mgas^{\prime}
\end{split}
\label{eqn:lnondet}
\end{equation}
describe the $n$ HI-detected galaxies and the $N{-}n$ non-detections, respectively.
In the above equations, $P(\lg\mgas \mid \ms, g{-}r, \bm{\theta})$ is the probability distribution of a galaxy
having a log-HI mass $\lg\mgas$ given its stellar mass and colour, and can be calculated as
\begin{equation}
    \begin{split}
    &P(\lg\mgas \mid \ms, g{-}r, \bm{\theta}) = \frac{1}{\sqrt{2\pi}\sig}\;\times\\
    &\exp\left\{-\frac{\left[\lg\mgas-\lg\ms-a \lg\ms - b (g{-}r) - c\right]^2}{2\sig^2}\right\}.
    \end{split}
\end{equation}
For computing the likelihood for the non-detections in Equation~\ref{eqn:lnondet}, I adopt an integration
limit of $\lg\mgas^{\mathrm{max}}{=}12$, consistent with the parameterisation of detection rate in Equation~\ref{eqn:fdet}.

\subsection{Posterior Results}
\label{subsec:results}

\begin{figure*}
\begin{center}
    \includegraphics[width=0.96\textwidth]{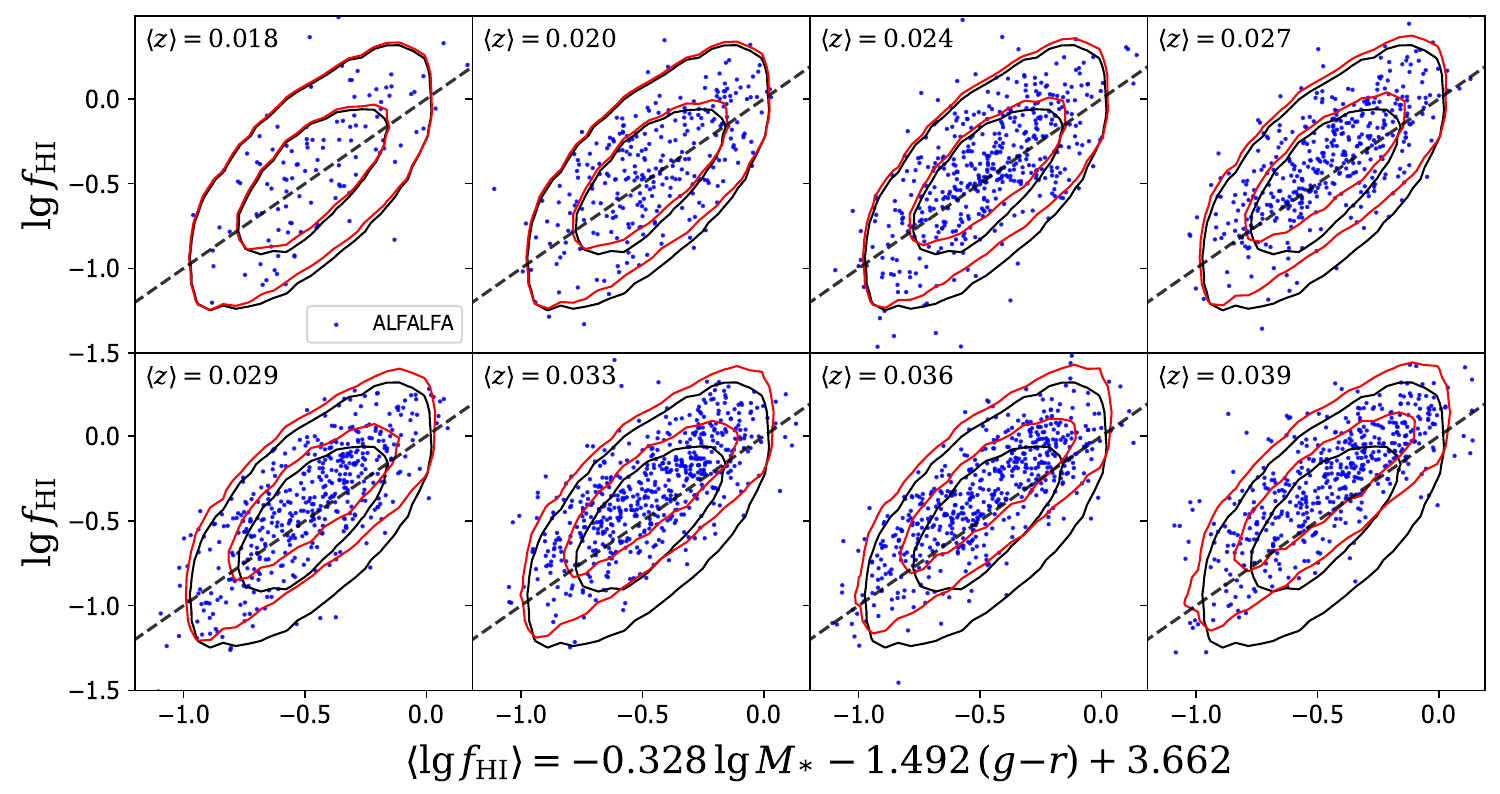}
    \caption{Comparison between the observed and
	predicted HI gas fractions for the joint SDSS-ALFALFA sample in eight
    different redshift slices. In each panel, black contours~(50\% and 90\%) indicate the
	predicted galaxy distribution on the $\fgas$ vs.
    $\langle\fgas\rangle$ plane of all the SDSS galaxies in the sample, while
    red contours show the predicted distribution of HI-detections by ALFALFA.
    Blue dots indicate the observed $\fgas$ in ALFALFA vs. their expected value
    $\langle\fgas\rangle$. The best-fitting HI fraction predictor for
    $\langle\fgas\rangle$ is shown at the bottom of the figure. The gray dashed
    line is the one-to-one line of $\fgas{=}\langle\fgas\rangle$.}
    \label{fig:FP_hiratio}
\end{center}
\end{figure*}

I apply the likelihood model to the SDSS-ALFALFA joint blue galaxy sample, and
infer the posterior distributions of the ten parameters using the Markov Chain
Monte-Carlo~(MCMC) method \texttt{emcee}~\citep{emcee}. After marginalizing over
the six nuisance parameters that describe the HI detection rate, I obtain the
posterior PDFs of the four key parameters that determine the HI fraction
predictor~($a,\,b,\,c$) and the log-normal scatter $\sig$, as shown in
Figure~\ref{fig:glory}.

\begin{figure*}
\begin{center}
    \includegraphics[width=0.88\textwidth]{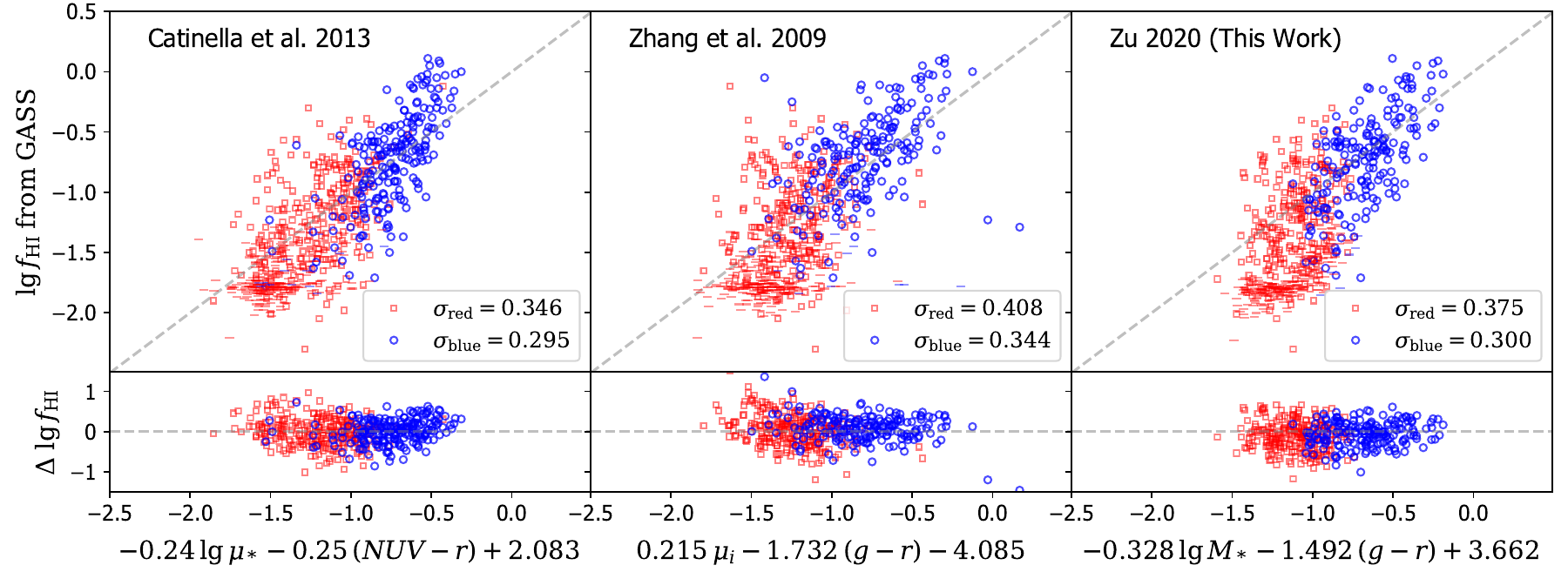} \caption{
        Comparison of the $f_{\mathrm{gas}}$ estimators derived by \citet{catinella2013}~(left),
        \citet{zhang2019}~(middle), and this work, using the
	$f_{\mathrm{gas}}$-limited data from the GASS reference sample. In each
	upper panel, blue circles and red squares indicate the HI-detected blue
	and red galaxies, respectively.
    The horizontal bars indicate the upper limits of $f_{\mathrm{gas}}$ for
    galaxies without robust HI detections. The scatters for the red and blue galaxies are
    shown by the legend in the bottom right. The differences between the observed and
    expected $f_{\mathrm{gas}}$ values are shown in the lower panels.
    }
\label{fig:gass}
\end{center}
\end{figure*}

The diagonal panels of Figure~\ref{fig:glory} show the marginalized 1D posterior
PDFs for each of the four key parameters, and the 1-$\sigma$ constraints are
$a{=}-0.328\pm0.015$, $b{=}-1.492\pm0.046$, $c{=}3.662\pm0.139$, and
$\sig{=}0.272\pm0.004$, respectively. In the off-diagonal panels, the red,
magenta,
and blue contour lines enclose the $68\%$, $95\%$, and $98\%$ confidence regions, respectively. The strong
correlation between $a$ and $c$ indicates that the model could potentially
explain away some of the apparent trend of $\fgas$ with $\ms$ by invoking a
strong Malmquist bias, but the fact that $P(a|\bm{\mathcal{D}})$ diminishes
rapidly to zero around $a{=}-0.28$ demonstrates that a negative intrinsic
correlation between $\fgas$ and $\ms$ at fixed $g{-}r$ is still necessary for
interpreting the data.

I simultaneously derive 1-$\sigma$ constraints on the six nuisance parameters
that describe the HI detection probability~(not shown on
Figure~\ref{fig:glory}), including $\lg M_{\mathrm{HI},0}{=}8.564\pm 0.054$,
$\mu_0{=}7.463\pm1.810$, $\lg M_{\mathrm{HI},1}{=}9.170\pm0.010$,
$\mu_1{=}14.386\pm0.920$, $\lg M_{\mathrm{HI},2}{=}9.427\pm0.013$, and
$\mu_2{=}30.822\pm2.270$.

From the best-fitting key parameters,
we can predict the underlying HI mass function of the volume-limited sample by
summing the probability distribution of HI mass of all galaxies in the sample.
The results for eight different redshift bins are shown as the black solid
curves in Figure~\ref{fig:detection_rate}. In comparison, blue histograms with
errorbars are the {\it observed} HI mass functions from ALFALFA. The ratios
between the histograms and the solid curves, i.e., the inferred detection rate,
are shown as the gray dashed lines~(with shaded uncertainty bands) in the bottom
sub-panels, while the blue solid curves are the predictions from the
best-fitting detection rate parameters.

\begin{figure*}
\begin{center}
    \includegraphics[width=0.96\textwidth]{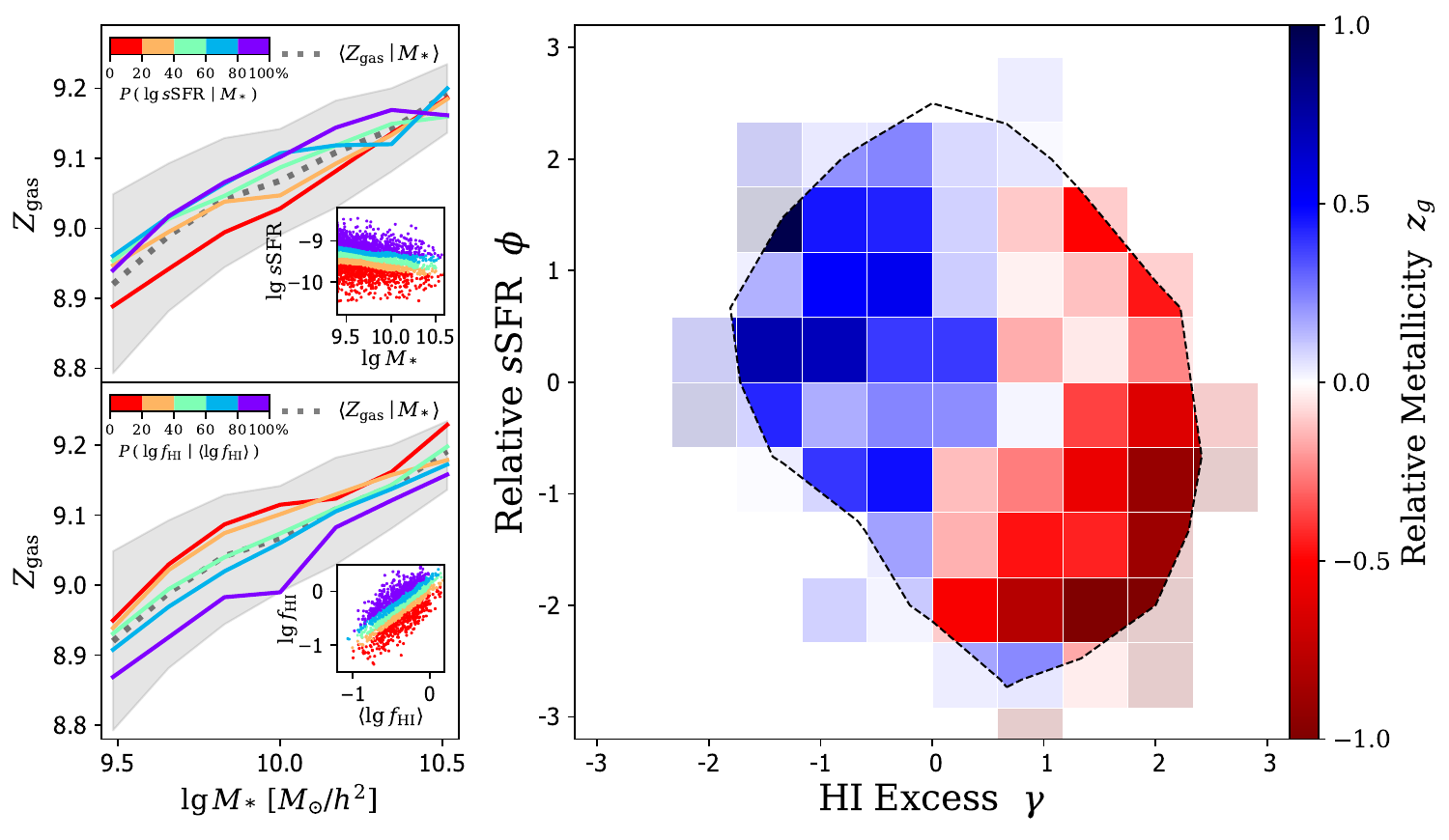} \caption{
	Top left: Dependence of the mass-metallicity relation~(MZR) on specific
	star formation rate. The inset
    panel shows the distribution of galaxies on the $\ssfr$-$M_*$ diagram, with
    each quintile colour-coded by the colour bar on the top left corner. The MZR
    of each quintile is indicated by the curve of the same colour in the main
    panel. The gray dotted curve and the shaded band indicate the median MZR and
    the scatter, respectively. Bottom left: Similar to the top left panel, but
    for the dependence of MZR on HI excess. Right: Distribution of the average
    relative metallicity $\zgas$~(normalized by the scatter in the MZR;
    indicated by the vertical colour bar on the right) on the relative
    $\ssfr$~($\phi$) vs. HI excess~($\gamma$) plane. The dashed contour
    highlights the high number density region that encloses
    $90\%$ of the sample. Clearly, the variation of relative metallicity $\zgas$ across the 2D plane is
    primarily driven by changes in the HI excess $\gamma$.}
\label{fig:mzr_3p}
\end{center}
\end{figure*}

At the lowest redshift~(top left of Figure~\ref{fig:detection_rate}; $\langle
z\rangle=0.018$), the observed HI source number counts are in good agreement
with the model prediction above $\lg\mgas{>}8.5$, indicating that ALFALFA is
capable of detecting most of the galaxies with $\lg\ms{>}9.4$ as HI sources. As
the redshift increases~(from left to right, top to bottom), the ALFALFA survey
missed progressively more and more Hi-rich systems --- at $z{=}0.04$~(bottom
right), ALFALFA only detected the most HI-rich systems and is highly incomplete
at $\lg\mgas{<}9.5$.

Figure~\ref{fig:FP_hiratio} provides a more visually-appealing way of showing
the impact of Malmquist bias on ALFALFA detections in eight different redshift
bins. In Figure~\ref{fig:FP_hiratio}, I compare the expected distribution that would be recovered when
comparing the observed $\fgas$ vs. the one predicted by my best-fitting
model based on a pure random distribution around the best-fitting estimator,
\begin{equation}
    \langle\lg\fgas\rangle = -0.328\lg\ms - 1.492(g{-}r) + 3.662,
    \label{eqn:hifpmean}
\end{equation}
with a scatter of $\sig{=}0.272$ dex. For doing so I create a Monte Carlo
simulation of $\fgas$ using as input the $M_*$ and g{-}r colours from my SDSS
sample, deriving the $\langle\lg\fgas\rangle$ based on
Equation~\ref{eqn:hifpmean}, and adding random noise. I repeat the process $100$ times
to recover the expected distribution, which is shown as the black contours in
each panel~($50\%$ and $90\%$). Then, for each of the eight redshift bins, I compare with the actual
distribution of observed $\fgas$ (detected by ALFALFA; blue dots) along the
expected distributions based on my best-fitting detection probability at that
redshift~(red contours). The difference between the black and red contours is
entirely due to the Malmquist bias.

In each panel, the ALFALFA-detected galaxies~(blue dots) show great agreement
with the model prediction~(red contours). At low redshifts~($z{<}0.02$), the red
contours are very similar compared to the black ones, but land primarily above
the one-to-one line~(dashed line) at high redshifts~($z{>}0.03$) --- ALFALFA
preferentially detected galaxies that have excess HI mass than expected. Without
correctly accounting for the Malmquist bias in the ALFALFA data, one would
derive an HI fraction predictor that systematically over-predicts the average HI
mass in galaxies~(by more than $0.15$ dex at $z{\sim}0.04$). Such Malmquist bias
correction is likely important for studies that utilize the HI mass function to
infer the HI-to-halo mass relation~(HIHM) using abundance matching
techniques~\citep{padmanabhan2017}.

One way to test the accuracy of my method is to observe the $\fgas$ for a
subsample of the galaxies that are undetected in ALFALFA, and compare to the
predictions by Equation~\ref{eqn:hifpmean}. The GASS Reference
Sample~\citep[DR3;][]{catinella2013} provides an ideal data set for such a test,
as include ${\sim}800$ galaxies that are observed to a limiting HI fraction of
$\fgas{=}2-5\%$, much lower than that of the ALFALFA survey. I apply my $\fgas$
predictor to the $\fgas$-limited the GASS Reference Sample, and compare my
estimator with two other similar methods from \citet{catinella2013}~(left) and
\citet{zhang2009}~(middle) in Figure~\ref{fig:gass}.  In each panel, the red
square and blue circles show the distributions of the red and blue
galaxies~(split by Equation\ref{eqn:cut}), respectively, on the observed vs.
predicted $\fgas$ plane. The scatters of the $\fgas$ estimator for red and blue
galaxies are listed by legend in the bottom right corner, while the horizontal
bars clustered
around $\fgas{\sim}2\%$ indicate the upper limits for galaxies without robust HI
detections. In the lower panels, I show the differences between the observed
and predicted $\fgas$ for the HI-detected systems. Similar to
Equation~\ref{eqn:hifpmean}, the \citet{catinella2013} and \cite{zhang2009}
estimators are both linear $\fgas$ predictors that rely on two photometric
observables. As indicated by the axis labels of the lower panels, the
\citet{catinella2013} estimator is the linear combination of the surface stellar
mass density $\mu_*$ and the NUV$-$r colour, while the \citet{zhang2009} method
relies on the surface brightness in $i$-band $\mu_i$ and the g$-$r color.  I do
not compare to linear models that utilize three or more
observables~\citep[e.g.,][]{li2012} or non-linear
models~\citep[e.g.,][]{teimoorinia2017, rafieferantsoa2018}, due to the
difficulty in evaluating their performances relative to mine.

The right panel of Figure~\ref{fig:gass} shows that my estimator provides a
great description of the GASS galaxies, regardless of their $\fgas$ or colour.
In particular, the scatter of my estimator for the blue galaxies~($0.3$ dex) is
comparable to the that of \citet{catinella2013}~($0.295$ dex), which is a direct
fit to the GASS data. Although Equation~\ref{eqn:hifpmean} is derived from blue
galaxies only, it still does a decent job describing the HI gas fraction of the
red galaxies, with a scatter~($0.375$ dex) only slightly higher than
\citet{catinella2013}~($0.346$ dex). Interestingly, the \citet{catinella2013}
estimator is biased high at both the low and high $\fgas$ ends, probably because
the blue and red galaxies do not exhibit the same HI gas dependences on the
surface mass density and NUV-r colour. In contrast, my estimator is unbiased
across all the entire range of $\fgas$. The \citet{zhang2009} estimator has
significantly larger scatters than the other two estimators, due to the presence
of several outliers. Overall, Figure~\ref{fig:gass} demonstrates the efficacy of
my estimator in mitigating the Malmquist bias and predicting the HI gas mass
for not only the high-$\fgas$ systems, but also for the enormous number of
low-$\fgas$ galaxies that largely evade the detections of modern HI surveys.

Finally, for any given galaxy with observed stellar mass and colour, I can now
predict the expected value of its HI mass fraction(Equation~\ref{eqn:hifpmean})
that is statistically consistent with the overall abundance of HI galaxies
detected and {\it missed} by ALFALFA. Beyond the mean HI fraction, the inferred
amount of scatter~($0.272$ dex) is dominated by the intrinsic scatter, as the
contribution from measurement uncertainties is very small~(${\sim}0.05$ dex).
This intrinsic scatter~($\sqrt{0.272^2-0.05^2}{=}0.267$ dex) was driven by a
myriad of physical processes involved in the build-up and depletion of
individual HI reservoir, which were inevitably linked to the galactic chemical
evolution and the large-scale density environment.  Therefore, I will explore
the connection between HI and the gas-phase metallicity and red galaxy
overdensity in the next two sections.

\section{What Drives the Mass-Metallicity Relation?}
\label{sec:mzr}

\subsection{Relative Metallicity, Relative s$\mathbf{SFR}$, and HI Excess}
\label{subsec:relative}

The mass-metallicity relation~(MZR) is a tight scaling relationship between the
stellar mass $\ms$ and the gas-phase metallicity $\zgas$ of star-forming
galaxies, with a scatter of ${\sim}0.1$ dex in the distribution of $\zgas$ at
fixed $\ms$. Defined as $\zgas{\equiv}12+\lg\,(\mathrm{O}/\mathrm{H})$, $\zgas$
is essentially a measure of the oxygen to hydrogen number density ratio in the
ISM. Therefore, at fixed $\ms$, a galaxy can be perturbed away from the median
MZR either by varying the oxygen abundance in the ISM via star formation and
galactic outflow, or by modifying the gas reservoir via gas inflow and
stripping.  To gain more insight on the physical driver of the MZR, I will
adopt this perturbative viewpoint and measure the variation of metallicity with
small changes in SFR and HI gas fraction.

The star-forming galaxies also form a relatively narrow sequence on the
$\lg\,\ssfr$-$\ms$ plane~(a.k.a., star-formation main sequence; SFMS).  I
define the mean MZR and SFMS by spline-interpolating the median $\zgas$ and
$\ssfr$ in small bins of stellar mass~($0.2$ dex), respectively. Taking
advantage of the MZR and SFMS, I can define a {\it relative} metallicity $z_g$
and a {\it relative} $\ssfr$ $\phi$ as the deviations of the observed
metallicity and logarithmic $\ssfr$ from the two mean scaling relations~(each
normalized by the scatter at fixed $\ms$), respectively.  Specifically, we
define $z_g$ and $\phi$ as follows,
\begin{equation}
z_g = \frac{\zgas - \langle Z_{\mathrm{gas}} \mid \ms \rangle}{\sigma_{\zgas\mid \ms}},
\end{equation}
and
\begin{equation}
    \phi = \frac{\lg\,\ssfr - \langle \lg\,\ssfr \mid \ms \rangle}{\sigma_{\lg\,\ssfr \mid \ms}},
\end{equation}
where $\langle Z_{\mathrm{gas}} \mid \ms \rangle$ and $\langle \lg\,\ssfr \mid
\ms \rangle$ are the mean MZR and SFMS, respectively. $\sigma_{\zgas\mid \ms}$
and $\sigma_{\lg\,\ssfr \mid \ms}$ are the two corresponding scatters at fixed
$\ms$. After switching to $z_g$ and $\phi$ defined above, I can now directly
compare the relative amount of metals and star formation of two arbitrary
galaxies, regardless of how different their stellar masses are.

Similarly, the HI fraction of star-forming galaxies follows a tight scaling
relationship about the best-fitting $\langle\fgas\rangle$
model~(Equation~\ref{eqn:hifpmean}) with a scatter of $\sig=0.272$ dex. To
quantify the tendency of a galaxy to have excess or deficit amount of HI
relative to its expected value, I define an ``HI excess'' parameter $\gamma$,
\begin{equation}
    \gamma = \frac{\lg {\fgas} - \lg {\langle\fgas\rangle}}{\sig},
\end{equation}
where $\fgas$ and $\langle\fgas\rangle$ are the observed and expected HI-to-stellar mass ratios, respectively.

\subsection{2D Relative Metallicity Map}
\label{subsec:mzr}

To study the 2D dependence of $z_g$ on $\phi$ and $\gamma$, I select from the
SDSS-ALFALFA joint sample a subset of HI-detected galaxies that have also high
S/N spectra observed by SDSS, so that each of those galaxy has all three
properties~(i.e., metallicity, $\ssfr$, and HI mass) robustly measured. I
eliminated the spurious metallicity measurements due to AGNs by imposing the BPT
selection criteria for star-forming galaxies defined by \citet{Kauffmann2003b}.
After the selections, I have $1{,}913$ galaxies in the subsample.

The left two panels of Figure~\ref{fig:mzr_3p} summarizes the individual
dependences of MZR on $\phi$~(top left) and $\gamma$~(bottom left),
respectively. In each panel, the gray dotted curve and shaded band indicate the
mean MZR and its scatter; The five coloured solid lines are the mean MZR of
galaxies in five quintiles of $\phi$~(top) or $\gamma$~(bottom), colour-coded by
the colour bar on the top left of each panel. The two inset panels show the
segregated distributions of galaxy quintiles on the SFMS diagram and the
$\lg\,\fgas$ vs.  $\langle\lg\,\fgas\rangle$ plane, respectively. For the MZR
dependence on $\phi$ shown in the top left panel, the lowest-$\phi$ quintile
exhibits the lowest average metallicity, while the metallicity trend with the
four higher-$\phi$ quintiles are less clear. The lowest-$\phi$ galaxies may have
insufficient star formation to chemically enrich the entire gas reservoir, or
preferentially live in systems with an excess of HI gas that dilutes the
metallicity. Conversely, metallicity and HI excess exhibit a strong
anti-correlation in the bottom left panel, where the five MZRs of the subsamples
form a well-defined decreasing sequence with increasing $\gamma$. This
anti-correlation can be naturally explained if HI excess is the underlying
driver of MZR.

The observed metallicity trend with $\phi$ indicates that there exists some weak
positive correlation between $\phi$ and $z_g$, in apparent contradiction with
the so-called ``fundamental metallicity relation''~(FMR), which states that the
gas-phase metallicity is anti-correlated with star formation rate. However, as
discussed in \S~\ref{sec:intro}, the existence of such an anti-correlation is
still under debate. In particular, the anti-correlation appears to be absent
from IFS observations~\citep{sanchez2013, sanchez2017, barreraballesteros2017,
sanchez2019}; And when detected in single-aperture observations, the shape and
amplitude of the correlation depends sensitively on the metallicity estimator
and galaxy selection~\citep{salim2014, kashino2016, telford2016}. For example,
using the same Bayesian metallicity estimator of~\citet{tremonti2004},
\citet{yates2012} found that the metallicity trend with SFR reverses its sign to
a positive correlation at the high stellar mass end from an anti-correlation at
the low mass, consistent with my finding in Figure~\ref{fig:mzr_3p}. However,
the trend disappears entirely if the metallicity estimator
of~\citet{mannucci2010} is used~\citep{salim2014}. Therefore, I emphasize that
despite being weaker than the metallicity trend with $\gamma$, the $\ssfr$ trend
seem in Figure~\ref{fig:mzr_3p} should be regarded as the maximal possible level
of correlation between $\ssfr$ and $z_g$ for the relevant stellar mass range.

To understand the strong metallicity trend with $\gamma$ and the lack of such a
trend with $\phi$, I measure the 2D relative metallicity distribution of
star-forming HI galaxies on the $\gamma$ vs. $\phi$ plane, as shown on the right
panel of Figure~\ref{fig:mzr_3p}. The colour of each pixel represents the
average relative metallicity $\langle z_g \rangle$ at given $\gamma$ and $\phi$,
colour-coded by the vertical colour bar on the
right. The dashed contour highlights the high number density region that encloses $90\%$ of the galaxies, revealing a weak anti-correlation between $\gamma$ and $\phi$.
One might worry that the aperture correction in the total $\ssfr$ might induce
spurious correlations between $\gamma$ and $\phi$, as the correction was based
on galaxy colors that I also employed to build the estimator for $\fgas$~(hence
$\phi$). To test the impact of such aperture bias, I calculate the Spearman's
cross-correlation coefficient $\rho_{cc}$ between $\phi$ and the aperture
correction of $\ssfr$, as defined by
$\lg\,\ssfr_{\mathrm{total}}/\ssfr_{\mathrm{fiber}}$. I do not find any
discernible correlation between the two quantities~($\rho_cc=-0.027\pm0.030$),
therefore the effect of aperture bias in $\ssfr$ can be safely ignored in
Figure~\ref{fig:mzr_3p}.

Clearly, the relative metallicity exhibits a strong dependence on HI excess, so
that galaxies with high $\gamma$ have significantly higher $z_g$ than those with
low $\gamma$, regardless of their relative $\ssfr$.  Conversely, the relative
metallicity shows little dependence on $\phi$ at fixed $\gamma$, suggesting that
SFR is a secondary driver of the gas-phase metallicity of a galaxy compared to
the amount of excess HI gas in that system. Therefore, the lack of a clear
metallicity trend with $\phi$ in the top left panel of Figure~\ref{fig:mzr_3p}
is expected, and the lowest metallicity exhibited by the lowest-$\phi$ quintile
can be entirely attributed to the anti-correlation between $\phi$ and $\gamma$,
which maps the lowest-$\phi$ quintile galaxies to the highest-$\gamma$~(hence
the lowest $z_g$) galaxies.

To summarize, my result in Figure~\ref{fig:mzr_3p} suggests that the scatter in
the MZR is primarily tied to the amount of excess HI gas in galaxies $\gamma$,
rather than the relative star formation rate $\phi$. It is commonly believed
that the MZR is mainly shaped by the balance between metal loss due to outflows
and metal production by stellar nucleosynthesis yield, both of which are tied to
star formation.  However, my result suggests that the dilution effect due to
inflows may have played a more important role in shaping the gas-phase
metallicity than the direct modification of metal abundance.  Therefore, we
emphasize that it is necessary to explicitly track the evolution of gas
reservoir in the analytic or semi-analytic exploration of the MZR.

\section{Environmental Dependence of HI}
\label{sec:env}

\subsection{Red Galaxy Overdensity}
\label{subsec:dred}

My HI fraction predictor is a function of stellar mass and optical colour,
without any direct dependence on the galaxy environment. Although stellar mass
and colour both depends strongly on the environment, the time scale over which
the environment modifies these two quantities is much longer than that of gas
stripping processes. In particular, the HI gas discs of satellite galaxies can
be rapidly stripped off by the ram pressure of the hot intra-cluster gas upon
infall, while their star formation activities can still last for a couple Gyrs
after infall~\citep{wetzel2013, simha2014}. For those satellite galaxies, there
would be no detectable change in the mass or colour of their stellar component
despite a sudden drop in the HI fraction~(see the paucity of HI-rich galaxies in
the \texttt{Coma} cluster in Figure~\ref{fig:wedge}).  Therefore, I expect that
the scatter in the HI fraction predictor is at least partly driven by the galaxy
environment.

\begin{figure}
\begin{center}
    \includegraphics[width=0.48\textwidth]{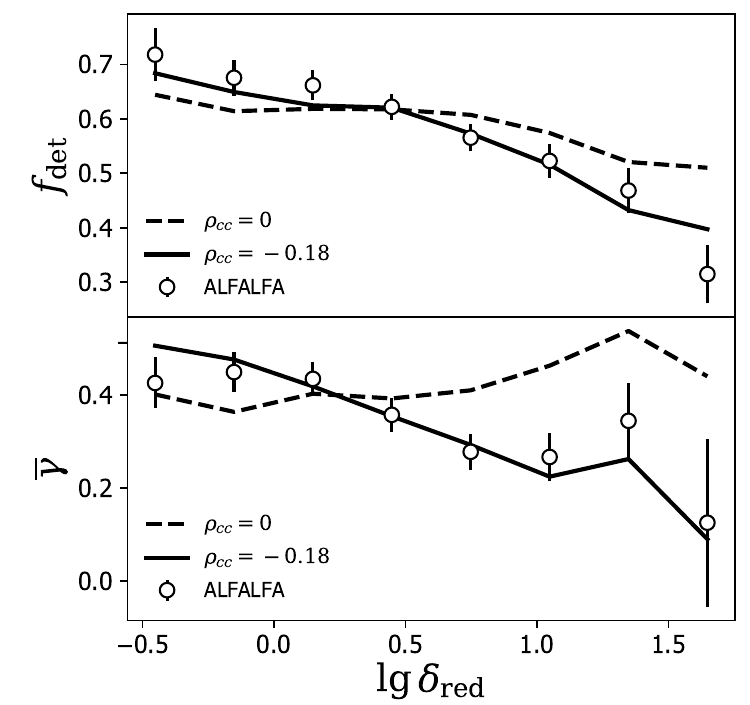} \caption{
	Top: HI detection rate as a function of $\delta_{\mathrm{red}}$. Open
	circles with errorbars show the HI detection rate in ALFALFA, while the
	dashed and solid lines are the predictions from two HI mocks that assume
	different cross-correlation coefficients between HI excess and
	$\delta_{\mathrm{red}}$, ${\rho_{cc}}=0$ and $-0.18$, respectively. A
	weak anti-correlation between HI excess and
	$\delta_{\mathrm{red}}$~(-0.18) is required to reproduce the observed
	trend of HI detection rate with $\delta_{\mathrm{red}}$. Bottom: Similar
	as above, but for the average HI excess as a
    function of $\delta_{\mathrm{red}}$. Again, the $\rho_{cc}=-0.18$ mock
    successfully reproduces the ALFALFA
	observation.}
\label{fig:delta_red}
\end{center}
\end{figure}

Ideally, I would prefer using halo mass as the main environment indicator, as ram pressure scales with the
product of the hot halo gas density and the infall velocity squared, both of which depend on halo mass.
Furthermore, the observed small scatter in the so-called ``Baryonic Tully-Fisher
Relation''~\citep{mcgaugh2000, lelli2016} implies that there exists a strong correlation between HI gas mass
and halo mass at fixed stellar mass. However, observationally I can only measure the average halo mass for an
ensemble of galaxies~\citep{mandelbaum2016}, but not yet for individual systems. Alternatively, one can employ
the empirical group catalogue and estimate halo masses from the abundance matching method~(AM), albeit with
significant scatter between the AM and true halo masses. Using the \citeauthor{yang2007} SDSS group catalog,
\citet{yoon2015} found strong radial variation of the HI detection fraction inside the most massive groups.
However, they failed to detect any dependence of HI fraction on the AM halo mass, likely due to the
combination of scatter and poor statistics (i.e., small number of rich groups).

\begin{figure*}
\begin{center}
    \includegraphics[width=0.96\textwidth]{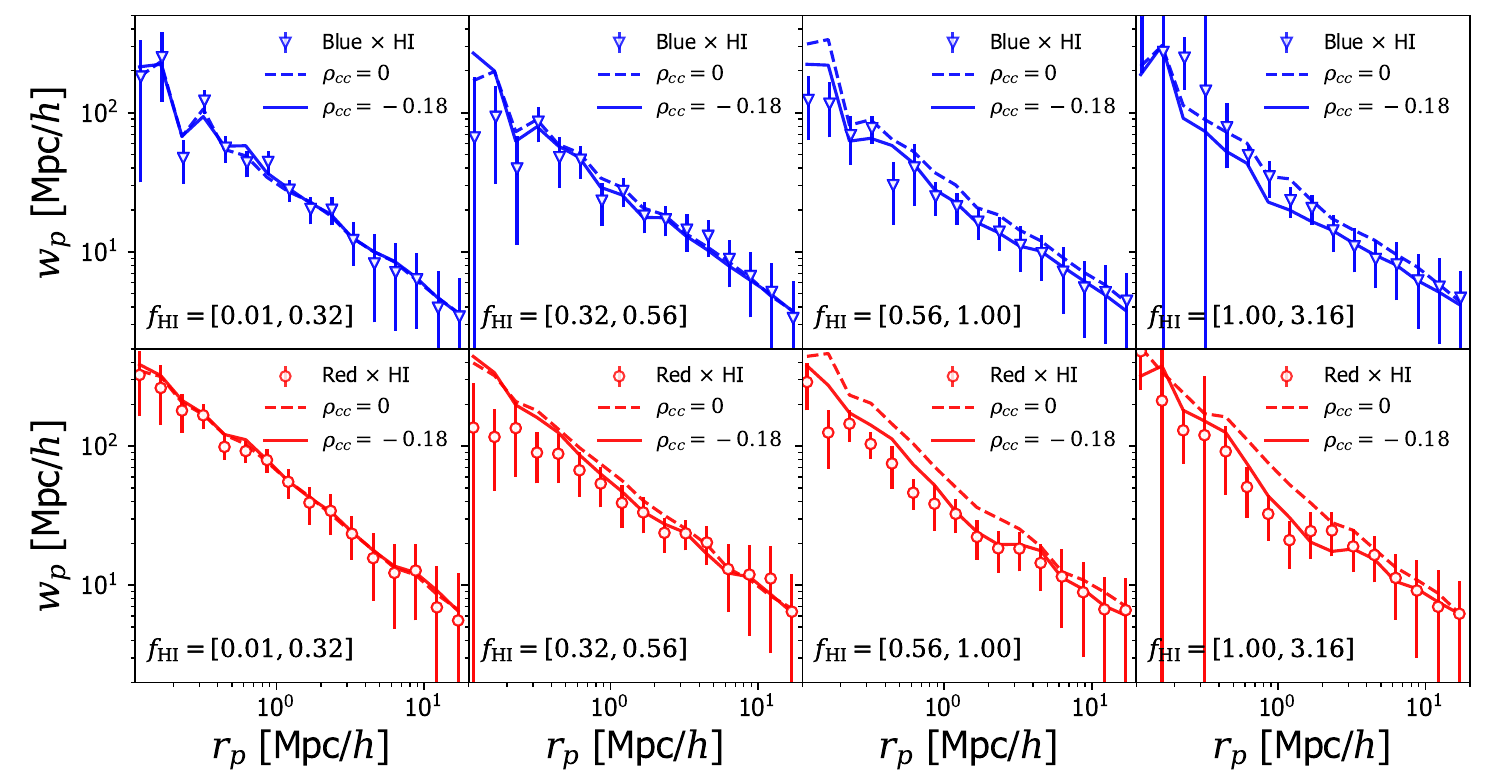} \caption{Projected cross-correlation
	functions $w_p$ between HI-detected galaxies and the red~(top row) and blue~(bottom row) galaxies in
	SDSS. In each row, the HI-detections are divided into four bins based on their HI gas fraction
	$\fgas$, from HI-poor~(left two panels; $[0.01, 0.32]$, $[0.32, 0.56]$) to HI-rich~(right two panels;
	$[0.56, 1]$, $[1, 3.16]$). In each panel, symbols with errorbars show the $w_p$ between SDSS red/blue
	galaxies and the ALFALFA detections, while the dashed and solid curves indicate the expected $w_p$
	from the two HI mocks with different cross-correlation coefficients between HI excess and
	$\delta_{\mathrm{red}}$, ${\rho_{cc}}=0$ and $-0.18$, respectively. The HI mock with a weak
    anti-correlation~($\rho_{cc}=-0.18$) successfully reproduces the observed projected correlation functions
    between red/blue and HI galaxies.}
\label{fig:wps_delta}
\end{center}
\end{figure*}

Inspired by the observed strong correlation between the red galaxy overdensity and halo mass~\citep{rozo2014,
    zu2016, zu2018}, I adopt the red galaxy overdensity $\dred$ around each blue galaxy
    as my proxy for
    galaxy
environment, defined as
\begin{equation}
    \dred = \frac{N_{\mathrm{red}}^{\mathrm{obs}}}{N_{\mathrm{red}}^{\mathrm{ran}}},
\end{equation}
where $N_{\mathrm{red}}^{\mathrm{obs}}$ and $N_{\mathrm{red}}^{\mathrm{ran}}$ are the number counts of
observed and random red galaxies within a cylindrical volume centered on that blue galaxy. The cylinder has an
aperture radius $R{=}2\hmpc$ and line-of-sight height $\Delta z{=}\pm 600\kms$. To correct for the survey
masks and boundary effect when computing $\dred$, I calculate $N_{\mathrm{red}}^{\mathrm{ran}}$ from a random
galaxy catalogue that has the same redshift and angular selection functions as the observed red galaxy sample.
The cylinder dimension is chosen roughly to match the projected radius and the
Fingers-of-God effect of massive cluster. I have verified that my conclusions are insensitive to the choice
of cylinder dimension.

If part of the scatter in $\fgas$ is driven by $\dred$, both the HI detection rate $\fdet$ and the average HI
excess $\overline{\gamma}$ of blue galaxies should depend on $\dred$, so that galaxies that live in
high-$\dred$ environments are more likely to be missed by ALFALFA and have a lower $\overline{\gamma}$ than
those in low-$\dred$ regions. As expected, the open circles in top and bottom panels of
Figure~\ref{fig:delta_red} show the observed declining trend of $\fdet$ and $\overline{\gamma}$ with
increasing $\dred$, respectively. The dashed lines are the predictions from the null assumption that there is
no dependence of $\gamma$ on $\dred$, so that the Spearman's rank correlation coefficient $\rho_{cc}$ between
$\gamma$ and $\dred$ is zero. In the $\rho_{cc}{=}0$ model, I randomly draw $\gamma$ values from a standard
normal distribution and assign mock $\mgas$ values to the blue galaxies using $\lg\mgas=\lg\ms + (\langle \lg
\fgas \rangle + \gamma \sig)$.  I then pass the mock $\mgas$ through the best-fitting detection probability
model at each redshift, thereby generating a mock SDSS-ALFALFA joint sample free of any environmental
dependences. Unsurprisingly, the dashed lines show no trend with $\dred$ in either panel, and the
$\rho_{cc}{=}0$ model is thus strongly disfavored by the ALFALFA observations.

To infer the cross-correlation coefficient $\rho_{cc}$ between $\gamma$ and $\dred$, one method is to find the
value of $\rho_{cc}$ that reproduces the observed $\fdet(\dred)$ in the top panel of
Figure~\ref{fig:delta_red}. Similarly, I generate the $\rho_{cc}{<}0$ mocks by imposing a negative Spearman's
rank correlation between $\gamma$ and $\dred$, while keeping the standard normal distribution of $\gamma$
intact. Using a simple minimum $\chi^2$ estimation that takes into account the Jackknife uncertainties of
$\fdet$, I find that the $\rho_{cc}{=}-0.18$ model~(solid line in the top panel) provides the best-fit to the
observed $\fdet(\dred)$.

A second method for inferring $\rho_{cc}$ is to find the best-fitting model that reproduces the observed
$\overline{\gamma}(\dred)$ in the bottom panel of Figure~\ref{fig:delta_red}. I repeat a similar $\chi^2$
fitting procedure using the $\overline{\gamma}(\dred)$ data, which also yield the best-fitting value of
$\rho_{cc}{=}=-0.18$~(solid line in the bottom panel). The excellent consistency between the two independent
methods of inferring $\rho_{cc}$ is highly non-trivial, as the first method relies more on the robustness of
my best-fitting detection probability model, whereas the second method depends more on the correctness of the
best-fitting HI fraction predictor. Therefore, this consistency not only confirms the existence of an
environmental dependence of HI in the local Universe, but also demonstrates the efficacy
of my comprehensive
model for interpreting the SDSS-ALFALFA joint dataset.

\subsection{Clustering Dependence on $\mathbf{f_{\mathbf{HI}}}$}
\label{subsec:wp}

For a more comprehensive study of the environmental dependence of HI, I measure the projected
cross-correlation functions $w_p$ between the HI-detected galaxies and the red vs. blue galaxies in SDSS.
The projected correlation function is computed as
\begin{equation}
    w_p(r_p) = \int_{-r_{\pi}^{\mathrm{max}}}^{r_{\pi}^{\mathrm{max}}}\;\xi^s(r_p,\,r_{\pi})\;\dd\,r_{\pi},
\end{equation}
where $r_p$ and $r_{\pi}$ are the projected and perpendicular distances between a pair of galaxies, and $\xi^s$
is the redshift-space cross-correlation function between HI and SDSS galaxies. I adopt an integration limit of
$r_{\pi}^{\mathrm{max}}{=}40\hmpc$ to reduce the impact due to peculiar motions. The 2D correlation function
$\xi^s$ is computed using the Davis \& Peebles estimator~\citep{davis1983},
\begin{equation}
\xi(r_p,\,r_{\pi}) = \frac{N_{R}}{N_{D}}\frac{HD}{HR} - 1,
\end{equation}
where HD is the number count of HI-SDSS galaxy pairs separated by ($r_p$, $r_{\pi}$), and HR is the number
count of pairs between an HI-detected galaxy and a point in the SDSS random catalogue. $N_{D}$ and $N_{R}$ are
the number of objects in the observed and random SDSS catalogues. I adopt the Davis \& Peebles estimator
because it only requires random catalogues for the SDSS samples, for which I have well-defined window
functions and masks~(but not for ALFALFA).  For the SDSS galaxies, I use two separate sets of randoms for the
blue and red galaxies, each with ten times the size of the observed sample.  I compute the measurement
uncertainties using 100 Jackknife subsamples defined over spatially contiguous patches on
the sky. I refer
readers to \citet{zu2015} for technical details in the construction of random catalogues and $w_p$
computation.

Similarly, I also compute $w_p$ between SDSS galaxies and the two mock HI samples constructed
in \S~\ref{subsec:dred}, i.e., mock ALFALFA observations assuming different levels of correlation between
$\dred$ and $\gamma$~($\rho_{cc}{=}0$ and $-0.18$). The $\rho_{cc}{=}-0.18$ mock has shown excellent agreement
with the observed $\fdet(\dred)$ and $\overline{\gamma}(\dred)$ in Figure~\ref{fig:delta_red}, but comparison
between the mock and SDSS-ALFALFA $w_p$ measurements would serve as a third and more stringent test of the
$\rho_{cc}{=}-0.18$ model --- $w_p$ directly measures the average density profile of red or blue galaxies
surrounding HI galaxies on small scales~(${<}4\hmpc$), as well as the clustering bias of HI galaxies on large
scales~(${\geq}4\hmpc$).

Figure~\ref{fig:wps_delta} shows the projected cross-correlation functions of blue~(top row) and red~(bottom
row) SDSS galaxies with HI-detected galaxies in four different bins of HI fraction $\fgas$~(increasing from
left to right). In the top~(bottom) row, blue triangles~(red circles) with errorbars are the cross-correlation
functions between HI and blue~(red) galaxies in the SDSS-ALFALFA joint sample. In each panel, dashed and solid
lines show the predictions from the $\rho_{cc}{=}0$ and $\rho_{cc}{=}-0.18$ HI mock samples, respectively.  In
the top panels, the mean density profile of blue galaxies around HI galaxies does not change with increasing
$\fgas$, while the large-scale clustering bias of HI galaxies increases slightly with $\fgas$. The
$\rho_{cc}{=}0$ mock~(dashed lines) provides reasonable description of the $w_p$ between blue galaxies and low
HI-fraction galaxies~($\fgas{<}0.56$), but over-predicts the amplitude of $w_p$ on all scales for high
HI-fraction systems~($\fgas{<}0.56$). In contrast, the $\rho_{cc}{=}-0.18$ measurements~(solid lines) of $w_p$
are in excellent agreement with the observed cross-correlations between ALFALFA and SDSS blue galaxies in all
four HI-fraction bins.

I examine the projected correlation of HI detections with the red galaxies in the bottom panels of
Figure~\ref{fig:wps_delta}. In contrast to the $w_p$ with blue galaxies, the mean density profile of red
galaxies around ALFALFA galaxies shows a strong suppression for those with $\fgas{>}0.56$ on scales
below $3\hmpc$~(compared to the dashed lines predicted by assuming $\rho_{cc}{=}0$), consistent with having an
environmental dependence of HI excess on $\dred$. This small-scale suppression of $w_p$ in the two
high-$\fgas$ bins is not captured by the $\rho_{cc}{=}0$ mock~(dashed lines), but the feature is successfully
reproduced by the $\rho_{cc}{=}-0.18$ mock~(solid lines), showing great consistency with the two tests in
Figure~\ref{fig:delta_red}. The significant lack of red galaxies in the vicinity of high-$\fgas$ systems
suggests that the massive halos play an important role in driving the HI-deficient systems in dense
environment, which would have otherwise been more HI-rich based on their stellar mass and
colour~(Equation~\ref{eqn:hifpmean}).

This environmental dependence of HI is likely induced by the various stripping
mechanisms in massive halos, including the tidal truncation of gas supply and
the ram-pressure stripping of gaseous discs, which usually require a careful
modelling of the dynamical interaction between galaxies and their host halos
using semi-analytic models or hydro-dynamic simulations.  However, we
demonstrate that the environmental dependence of HI can be fully accounted for
by assuming a cross-correlation between the HI excess $\gamma$ and the red
galaxy overdensity $\dred$, and the cross-correlation coefficient $\rho_{cc}$
can be self-consistently inferred from the data.  Within the SDSS-ALFALFA joint
sample, I discover that there exists a weak negative correlation between
$\gamma$ and $\dred$, $\rho_{cc}{=}-0.18$, derived from three independent tests
illustrated in Figure~\ref{fig:delta_red} and Figure~\ref{fig:wps_delta}. I
note that this inferred correlation between $\gamma$ and $\dred$ can also be
used as an interesting constraint of other models of galaxy formation.

\section{Conclusion}
\label{sec:conc}

In this paper, I develop a statistical method to infer the HI-to-stellar mass ratio $\fgas$ of galaxies from
their stellar mass and optical colour, using a volume-limited galaxy sample jointly observed by SDSS and
ALFALFA. Compared to the traditional methods, the key feature of my method is its capability of removing the
Malmquist bias against low-$\fgas$ systems in ALFALFA, via a self-consistent modelling of the HI detection
rate of each galaxy observed in SDSS. The best-fitting HI fraction predictor has an estimated scatter of
$0.272$ dex, slightly smaller than the ${\sim}0.30$ dex reported by traditional methods.

To explore the impact of gas accretion on gas-phase metallicity, I define an HI
excess parameter $\gamma$ as the deviation of the observed $\lg\,\fgas$ from the
expected value (normalized by scatter). I discover a strong secondary dependence
of the mass-metallicity relation on $\gamma$, echoing the findings of
\citet{bothwell2013} and \citet{brown2018}. This secondary dependence defines a
fundamental metallicity relation of HI, similar to the fundamental metallicity
relation of the star formation rate~\citep{mannucci2010, laralopez2010,
andrews2013}.

By taking advantage of two tight scaling relations, i.e., the mass-metallicity
relation and the star formation main sequence, I define the relative metallicity
and relative $\ssfr$ of each galaxy as the (normalized) deviations from the two
respective mean relations.  To elucidate the underlying driver of the scatter in
the MZR, I examine the 2D relative metallicity distribution on the relative
$\ssfr$ vs HI excess plane. I find that the variation of relative metallicity is
primarily driven by the change in HI excess, so that galaxies with higher HI
excesses always have lower relative metallicities, regardless of the difference
in relative $\ssfr$. This 2D metallicity map suggests that the
metallicity dependence on HI is more fundamental that on SFR.

Furthermore, the HI excess also depends on the large-scale overdensity
environmental. Using the red galaxy overdensity $\dred$ as a measure of the
large-scale environment, I demonstrate that there exists a weak
anti-correlation between HI excess and $\dred$ in the SDSS-ALFALFA joint sample.
From the dependence of detection rate and HI excess on $\dred$, I infer the
cross-correlation coefficient $\rho_{cc}$ between the two quantities to be
$-0.18$. The $\rho_{cc}{=}-0.18$ model also successfully reproduces the
dependence of HI clustering on $\fgas$. I believe this anti-correlation can be
largely explained by the ram pressure and tidal stripping of HI gas discs in
cluster environments~\citep[but see][]{wang2018}.

Currently, ALFALFA is the only extragalactic HI survey of a cosmological
volume~\citep{avila2018}. However, with the advent of exciting HI surveys like
the Square Kilometer Array~\citep[SKA;][]{maartens2015} and the
Five-hundred-meter Aperture Spherical Telescope~\citep[FAST;][]{nan2011}, the HI
sky will be observed to a much higher depth within a significantly larger volume
than ALFALFA. In particular, FAST will conduct the Commensal Radio Astronomy
FasT Survey~\citep[CRAFTS]{zhang2019}, a drift scan survey of ${\sim}600000$ HI
sources across $20000$ deg$^2$ sky up to $z{=}0.35$. My method will provide a
viable path to the synergy between the next-generation HI surveys like CRAFTS
and upcoming optical surveys, e.g., the Bright Galaxy Survey program within the
Dark Energy Spectroscopic Instrument~\cite[DESI;][]{desi2016}. In particular, we
expect the method to provide valuable insight into the evolution of HI gas and
metallicity in cluster environments~\citep{peng2014, li2018} and the dependence
of HI on large-scale tidal environments~\citep{liao2018, alam2018}.

\section*{Acknowledgements}

I thank the anonymous referee for suggestions that greatly improved the
manuscript, and Alessandro Sonnenfeld for helpful discussions. YZ acknowledges
the support by the National Key Basic Research and Development Program of China
(No.  2018YFA0404504), National Science Foundation of China (11621303,
11873038), the National One-Thousand Youth Talent Program of China, and the STJU
start-up fund (No. WF220407220).



\bibliographystyle{mnras}



\bsp	
\label{lastpage}

\end{document}